\documentclass[iop.revtex4]{emulateapj}
\usepackage{amsmath}
\def\kms{\rm{km \ s^{-1}}}
\def\etal{\rm{et al. }}
\def\deg{^\circ }

\tighten

\clearpage

\begin{document}

\title{Integral-Field Stellar and Ionized Gas Kinematics of Peculiar Virgo Cluster Spiral Galaxies}
\author{Juan R. Cort\'es\altaffilmark{1}
}
\affil{National Radio Astronomy Observatory}
\affil{Avenida Nueva Costanera 4091, Vitacura, Santiago, Chile}
\email{jcortes@alma.cl}
\affil{Joint ALMA Observatory}
\affil{Alonso de C\'ordova 3107, Vitacura, Santiago, Chile}
\affil{Departamento de Astronom\'{\i}a, Universidad de Chile}
\affil{Casilla 36-D, Santiago, Chile}
\author{Jeffrey D. P. Kenney}
\affil{Department of Astronomy, Yale University}
\affil{P.0. Box 208101, New Haven, CT 06520-8101}
\email{jeff.kenney@yale.edu}
\author{Eduardo Hardy\altaffilmark{1,2}
}
\affil{National Radio Astronomy Observatory}
\affil{Avenida Nueva Costanera 4091, Vitacura, Santiago, Chile}
\email{ehardy@nrao.cl}
\affil{Departamento de Astronom\'{\i}a, Universidad de Chile}
\affil{Casilla 36-D, Santiago, Chile}

\altaffiltext{1}{The National Radio Astronomy Observatory is a facility of the National
Science Foundation operated under cooperative agreement by Associated
Universities, Inc.}
\altaffiltext{2}{Adjoint Professor}

\received{someday}

\shorttitle{Some title}
\shortauthors{Cort\'es, Kenney \& Hardy}

\begin{abstract}
We present the stellar and ionized gas kinematics of 13 bright peculiar Virgo cluster galaxies observed with the DensePak Integral Field Unit at the WIYN 3.5-meter telescope, to seek kinematic evidence that these galaxies have experienced gravitational interactions or gas stripping. 2-Dimensional maps of the stellar velocity $V$, and stellar velocity dispersion $\sigma$ and the ionized gas velocity (H$\beta$ and/or [\ion{O}{3}])
are presented for galaxies in the sample.
The stellar rotation curves and velocity dispersion profiles are determined for 13 galaxies, and the ionized gas rotation curves are determined for 6 galaxies.
Misalignments between the optical and kinematical major axis are found in several galaxies. While in some cases this is due to a bar, in other cases it seems associated with a gravitational interaction or ongoing ram pressure stripping. Non-circular gas motions are found in nine galaxies, with various causes including bars, nuclear outflows, or gravitational disturbances. 
Several galaxies have signatures of kinematically distinct stellar components,
which are likely signatures of accretion or mergers.
We compute for all galaxies the angular momentum parameter $\lambda_{\rm R}$. An evaluation of the galaxies in the $\lambda_{\rm R}$-ellipticity plane shows that all but 2 of the galaxies have significant support from random stellar motions, and have likely experienced gravitational interactions.
This includes some galaxies with very small bulges and truncated/compact H$\alpha$ morphologies, indicating that such galaxies cannot be fully explained by simple ram pressure stripping, but must have had significant gravitational encounters.  Most of the sample galaxies show evidence for ICM-ISM stripping as well as gravitational interactions, indicating that the evolution of a significant fraction of cluster galaxies is likely strongly impacted by both effects.

\end{abstract}


\clearpage
\newpage
\clearpage

\newpage
\section{Introduction}

It is well known that the environment affects the morphological types of galaxies in clusters. 
Many studies show that galaxies in clusters evolve morphologically, 
with spirals becoming redder and in some cases lenticular as the result of environmental effects
(Dressler 1980; Butcher \& Oemler 1978, 1984; Dressler \etal 1997; Poggianti \etal 1999 ; Poggianti \etal 2009 ;
Kormedy \& Bender 2012)
Several mechanisms have been proposed for driving galaxy evolution,
including processes that affect the stars, gas, and dark matter, and those that affect only the gas.
In the first category we list
(i) low-velocity tidal interactions and mergers (e.g.;  (Toomre \& Toomre 1972; Hernquist 1992), 
(ii) high-velocity tidal interactions and collisions (e.g., Moore \etal 1996),
(iii) and tidal interaction between galaxies and the cluster as a whole or between galaxies and substructures within the cluster (Bekki 1999).
In the second category we list  
(i) Intracluster medium - interstellar medium (ICM-ISM) stripping
(Gunn \& Gott 1972; Nulsen 1982; Schulz \& Struck 2001; Vollmer
\etal 2001; van Gorkom 2004 ; Cen 2014), (ii) gas accretion, which may occur in
the outskirt of clusters, and (iii) starvation or strangulation, where
the galaxies could lose their gas reservoir thus preventing their accretion
onto the galaxy (Larson, Tinsley, \& Caldwell 1980).
While all of these processes probably do actually occur, it remains unclear 
which ones are dominant in driving the morphological evolution of cluster galaxies.

Detailed studies of the stellar and ionized gas kinematics can help
to discriminate between the different interaction processes. 
For example, gravitational interactions produce disturbed kinematics in both the stellar and
gas components, whereas  interactions of a hydrodynamic nature will directly affect only the gas.
Recently, with the advent of Integral Field Units (IFUs) such as DensePak,
SAURON, GMOS, SINFONI, and MUSE, these detailed studies become possible. 
The observed velocity fields can be compared with those from simulations
(e.g. Bendo \& Barnes 2000; Jesseit \etal 2007;  Kronberger \etal 2007; Kronberger \etal 2008), 
providing important clues about the physical processes that drive galaxy evolution.

The Virgo cluster is the nearest moderately rich cluster
with a galaxy population spanning a large range of
morphological types. The cluster has a moderately dense ICM, and is dynamically
young with on-going sub-cluster mergers and infalling galaxies,
making it into an ideal place
for detailed studies of various environmental processes. Moreover,
the Virgo cluster has a significant population
of galaxies characterized by truncated star formation morphologies,
with no H$\alpha$ in the outer disk but strong H$\alpha$ in the inner
region (Koopmann \& Kenney, 2004) consistent with ICM-ISM stripping.
However, some of them have in addition other peculiarities that are not
presently well understood,
presumably reflecting different types of interactions. These
peculiar galaxies may be in the process of morphological transformation,
and could be considered as ``snapshots'' in the evolutionary path from
actively star-forming spiral galaxies to more passive spirals and lenticulars.

With these objectives in mind we present in this work a study of the stellar
and ionized gas kinematics of 13 peculiar Virgo cluster galaxies using
integral-field spectroscopy techniques. 
We profit from the ability of this technique to accurately map two-dimensional velocity fields 
for both the stars and the ionized gas in the centers of these galaxies.
This data set has been previously used to estimate 
the 3D cluster location of all sample galaxies by using stellar kinematics to derive their distances
(Cort\'es \etal 2008), and for a detailed investigation of the nature of two of the most peculiar galaxies of the present sample,
NGC 4064 and NGC~4424 (e.g.; Cort\'es \etal 2006).

The present paper is structured as follow; A brief description of the galaxy sample is given in \S 2.
The observation and data reduction procedures are summarized in \S 3.
A description of the observed stellar and ionized gas kinematics is presented
in \S 4. In \S 5, we summarize the
kinematical peculiarities observed in these galaxies. In \S 6,
we discuss the velocity dispersion profiles and kinematical support of these galaxies. In \S 7,
we compare our observations with simulations of merger remnants, ICM-ISM stripped
galaxies, and tidal interactions.
These results are discussed from the perspective of galaxy evolution
in clusters in \S 8. We summarize our results and present our
conclusions in \S 9.
A discussion of individual galaxies is given in the Appendix.

\section{The Galaxy Sample}

\begin{figure}
\plotone{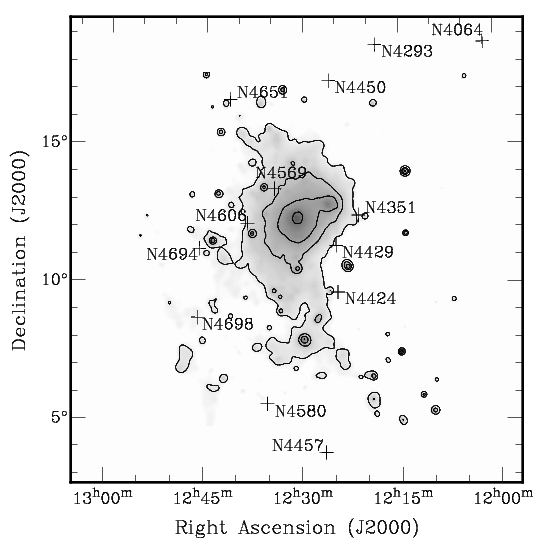}
\caption{Sample galaxies and their location in the Virgo Cluster. The contour map
represents the {\em  ROSAT} X-ray emission in the cluster
(B\"ohringer et al. 1994). Sample galaxies are represented by a black plus sign and its
NGC name.
 }
\label{galaxypos}
\end{figure}

The sample consists of 13 peculiar Virgo cluster spiral galaxies,
spanning a variety of optical
morphologies (Table $\ref{table1}$, Figure $\ref{galaxypos}$).
Morphological selection was made using the R and H$\alpha$ atlas of
Virgo cluster galaxies of
Koopmann \etal (2001), whereas the kinematical selection
made use of the published
H$\alpha$ kinematics on 89 Virgo cluster spirals by Rubin \etal (1999).
While the sample selection is not uniform, it is
designed to include bright Virgo spirals
whose peculiarities are most poorly understood, and to
include representatives of the different H$\alpha$ types identified
by Koopmann \& Kenney (2004).
In choosing sample galaxies within a given H$\alpha$ type,
we gave preference to those with kinematical peculiarities.

In Virgo spirals, there is a poor correlation between 
the bulge-to-disk ratio and the normalized star formation rate 
so that the Hubble classification of spirals does not work well in the Virgo Cluster
(Koopmann \& Kenney 1998).
The Hubble type classification assigned to Virgo galaxies generally reflects 
the star formation rate rather than the bulge-to-disk ratio, so that 
Virgo spirals with reduced star formation rates are generally classified as
early type spirals, independent of their bulge-to-disk ratio.
Our sample includes more early type than late type galaxies, since most of
the strongly disturbed cluster galaxies have reduced star formation rates and so are
classified as early types.
Since the Hubble classifications of cluster spirals do not capture the intrinsic variation in 
galaxy morphologies, either the distributions of old stellar light or the young stars,
we use instead the light concentration parameter C$_{\rm 30}$ (Abraham \etal 1994) as an objective measure of the
bulge-to-disk ratio, and the H$\alpha$ type (Koopmann \& Kenney 2004), 
to describe the radial distribution of star formation.

Since H$\alpha$ types seem to correlate with the type of interaction experienced by the galaxy
(Koopmann \& Kenney 2004),  and also to some kinematical properties that we describe in this paper,
here we give brief definitions of these categories, and indicate in which category our sample galaxies belong.

\begin{itemize}

\item {\em Normal:} NGC 4651.
In Normal galaxies, the H$\alpha$ radial distribution close to the mean of isolated spirals, both in 
the shape of the radial distribution (which is close to the R light profile), and in amplitude.

\item {\em Truncated/Normal:} NGC 4351, NGC 4457, NGC 4569, and NGC 4580.
In Truncated/Normal galaxies , the
H$\alpha$ radial distribution like that in a Normal galaxy out to a well-defined truncation radius, but 
there is virtually no H$\alpha$ emission beyond.

\item {\em Truncated/Compact:} NGC 4064, NGC 4424, NGC 4606, and NGC 4694.
In Truncated/Compact galaxies, the H$\alpha$ radial profile is much steeper than the R light profile at all radii,
with a strong central peak, and a very sharp drop with radius, such that there is virtually no emission beyond the
central ~1 kpc.   In Koopmann \& Kenney (2004) this category also included the provision that
the central H$\alpha$ intensity was much higher than Normal, but here we relax this provision.
Thus we include NGC 4694\footnotemark[1]\footnotetext[1]{In Koopmann \& Kenney 2004 NGC 4694 was classified as T/N,
but a re-inspection or the H$\alpha$ radial profile shows that it does not match the T/N criteria, and is better
described by the modified T/C definition proposed here.}
 in this category, which has an H$\alpha$ radial profile like the other
T/C galaxies, but with a lower amplitude.

\item {\em Anemic or Truncated/Anemic galaxies:}  NGC 4293, NGC 4429, NGC 4698, and NGC 4450.
In Anemic galaxies the
shape of the H$\alpha$ radial distribution is like that of a Normal galaxy (and like the R light profile), but
with a much lower amplitude. Anemic galaxies have weak but detectable emission over much of the stellar disk.
Truncated/Anemic galaxies have an anemic distribution out to a well-defined truncation radius, but 
virtually no H$\alpha$ emission beyond.
Whereas most of our sample galaxies in these categories have H$\alpha$ emission
detected out to at least ~0.3R$_{25}$,
we also include in the Truncated/Anemic category the S0 galaxy NGC 4429, 
which has a very small and very weak H$\alpha$ disk.

\end{itemize}

Distances to many individual early type Virgo galaxies have now been measured to good accuracy,
based on the surface brightness fluctuation method.
The mean distance to early type Virgo cluster galaxies is 16.5 $\pm$ 0.1 Mpc (Mei \etal 2007).
The distances to late type and peculiar Virgo galaxies are less accurately known.
The commonly used Tully-Fisher method based on HI line-widths works well for most late type galaxies,
but gives erroneous distances for a subset of cluster galaxies which are HI-poor and
disturbed (Cort\'es \etal 2008). In Cort\'es \etal (2008) we derived distances to galaxies in the present
Virgo sample based on a new approach using a stellar kinematics-based version of the the
Tully-Fisher relation. It is revealing to compare different distance estimates for perhaps
the most disturbed galaxy in our sample, NGC 4424. The HI-based Tully-Fisher method gives a distance
of 4.8 Mpc (Solanes \etal 2002),  our stellar-kinematics-based Tully-Fisher relation gives a distance
of 15.2 Mpc (Cort\'es \etal 2008), and a recent Type Ia supernova gives a distance of 15.5 Mpc
(Munari \etal 2013). While this is an extreme example, it illustrates that our stellar-kinematics-based
Tully-Fisher distances may be relatively accurate. We adopt these distances in this paper, which are given in Table $\ref{table1}$.

\section{Observations, Data Reduction, and Methods}

\subsection{Observations and Data Reduction}

The galaxies were observed by using the DensePak Integral-field unit
(Barden, Sawyer \& Honeycutt 1998)
installed
at the 3.5m WIYN telescope at Kitt Peak. The DensePak array consists of 90
fibers with a diameter of 3.5", and about 4" separation with a covering area
of 30"$\times$45", which corresponds to 2.3$\times$3.5 kpc at the mean Virgo
distance. The observations
were taken in three observing runs during April 8-9$^{th}$ 1999,
May 24-25$^{th}$ 2001,
and February 10-12$^{th}$ 2002. The 860 line mm$^{-1}$, blaze angle
30$\deg$.9 grating at 5000 {\AA} was used
at second
order, covering the 4500--5500 {\AA} wavelength range with a spectral
dispersion of 0.48 {\AA} per pixel, and a spectral resolution of 2.02 {\AA}.
The DensePak array, in most of the cases,
was oriented along the optical major axis of the galaxies
(Table $\ref{table2}$, Fig. $\ref{galaxydspk1}$).
Details about the exposure time used to observe the galaxies are given in
Table $\ref{table2}$.
Radial velocity standards stars (Barbier--Brosat \& Figon 2000)
were observed as spectral templates,
with an exposure time of 180 seconds. Details about each star are
presented in Table $\ref{table3}$.

The spectra were reduced in the usual way of IRAF reduction for multi--fiber
spectrographs. They were zero-subtracted and overscan corrected with the
standard IRAF tasks. The flat fielding correction, sky subtraction,
fiber throughput correction, and wavelength calibration were carried
out using the IRAF task
DOHYDRA in the package HYDRA (Vald\'es 1995).
Alignment between different exposures was checked by comparing the
continuum maps between each exposure. The exposures were aligned using
our own program written in IDL\footnotemark[1]\footnotetext[1]{www.exelisvis.com/ProductsServices/IDL.aspx}
After the alignment, the galaxy spectra were averaged for
improving signal-to-noise ratio and removing cosmic rays.

In two sample galaxies; NGC 4424 and NGC 4351, the spectra in many positions
has a signal-to-noise ratio per pixel lower than limit for obtaining reliable
kinematics until the second moment (SNR $\sim$ 15). Therefore, spectra in these
galaxies were binned using the adaptive scheme developed by
Cappellari \& Copin (2003), for achieving a signal-to-noise ratio good enough
for obtaining reliable kinematics.

\subsection{Stellar and Gas Kinematics}

Stellar kinematics of the sample galaxies, were derived by using the
penalized pixel-fitting (pPXF) method developed by
Cappellari \& Emsellem (2004). This
method allows the masking of emission lines, a more realistic
estimation of errors than other methods, and automation, allowing us to
derive the kinematics of many spectra in a relatively short amount of time.
Although, the method can measure the Gauss-Hermite moments of the
line-of-sight velocity distribution (LOSVD) up
to $h_{6}$, the low signal-to-noise ratio
of the spectra (SNR $<$ 60) in most of our sample galaxies, only allow us to obtain reliable measurements
until the velocity dispersion ($\sigma$). We obtained Gauss-Hermite moments up
to $h_{4}$, only in 3 sample galaxies; NGC 4429, NGC 4450, and NGC 4561.
Template spectra were chosen
from the spectra of observed radial velocity standard stars
(typically G \& K giants) which better fit the galaxy spectra.
The pPXF method determines the broadening function (LOSVD) between the template star spectra and the galaxy spectra.
If the instrumental dispersions between the template
and the galaxy are different, we need to convolve the template spectra by the square root of the
quadratic difference between the instrumental dispersion of the galaxy and the template star. Since, we observed the template
stars and sample galaxies using the same observational setup during the same observing runs,
so the instrumental
dispersions are the same for both, it was not necessary to convolve the template star, and the velocity dispersion derived is already corrected by instrumental dispersion.
A linear
combination of different template star didn't show any significant
improvement in the fits. The heliocentric correction was applied to the
derived
line-of-sight velocities in each galaxy. Errors were estimated by using a
Monte-Carlo scheme. They were obtained as the standard deviation of the
kinematical parameters ($V$ and $\sigma$) from
many realizations (N=100) of the
input spectra by adding Gaussian noise to a model galaxy spectrum.
Comparison between the galaxy spectra in the central region,
and the model galaxy spectra obtained by pPXF from the template star
spectra are shown in Fig. $\ref{samplespectra}$.

Two dimensional ionized gas kinematics were derived from the emission lines
(H$\beta$ and [\ion{O}{3}]$\lambda$5007) by fitting a Gaussian
function to the emission lines
in the continuum subtracted spectra, obtaining $V$ and $\sigma$ for the
ionized gas. Finally we applied the heliocentric correction to the derived gas velocities.
In this work, we focus only on the ionized gas
line-of-sight velocity.

\subsection{Kinematic parameter determination}

Systemic velocities, kinematic position angles, and rotational velocities
were obtained for each galaxy by fitting a
pure circular tilted ring model (Begeman 1989) to the stellar and
ionized gas velocity fields, in cases where this was possible.

The method consists of dividing the galaxy into a set of concentric rings, each
ring being characterised by a fixed value of the inclination $i$, the rotation velocity $V_{\rm rot}$ and
kinematic position angle $\phi$. We define the sky coordinates $(x,y)$ as
the DensePak array coordinates with $x$ oriented over the position angle of the
DensePak array $\phi_{\rm DensePak}$ (see Table $\ref{table2}$). For a given ring, the observed radial velocities recorded
on a set of sky-coordinates are given by;

\begin{equation}
V_{\rm obs}(x,y)=V_{\rm sys}+V_{\rm rot}(R) \sin(i) \cos(\theta),
\end{equation}
where $\theta$ is the azimuthal angle in the plane of the galaxy, measured from the optical major
axis of the galaxy. It can be shown that for any point $(x_{\rm gal},y_{\rm gal})$ in the plane of
the galaxy, $\theta$ is given by;

\begin{equation}
\begin{split}
cos(\theta) = & \frac{x_{\rm gal}}{R} = \frac{1}{R} [ (x -x_{0}) \cos (\phi - \phi_{\rm DensePak}) \\
             & + (y-y_{0}) \sin(\phi - \phi_{\rm DensePak}) ],
\end{split}
\end{equation}
\begin{equation}
\begin{split}
sin(\theta)= & \frac{y_{\rm gal}}{R}= \frac{1}{R \cos(i)} [(y-y_{0}) \cos(\phi - \phi_{\rm DensePak}) \\
            &  - (x-x_{0}) \sin (\phi - \phi_{\rm DensePak}) ],
\end{split}
\end{equation}
where $R$ is the mean radius of the ring in the plane of the galaxy, and $(x_{0},y_{0})$ are the sky
coordinates of the center of the ring.
In this work, we define position angle (either optical or kinematic) as the angle between
the optical/kinematic major axis of the galaxy, and the line from the galaxy center headed North, measured from the North to East.

The procedure is iterative, and described as follows:
1) We divide the galaxy into concentric rings with typical widths of one fiber on the major axis. 2) For each
ring, we make a least-square fit of $V_{\rm obs}$ with  $x_{0},y_{0}$ and $V_{sys}$ as free parameters, keeping fixed $V_{\rm rot}$, $\phi$ and
$i$. Then, we took as center coordinates and systemic velocity the mean values over the all the rings. 3) After this,
we kept fixed  $x_{0},y_{0}$ and $V_{\rm sys}$, and we fitted the values $V_{\rm rot}$, and $\phi$, simultaneously.
We didn't attempt to fit the inclination $i$ as it was difficult to disentangle the degeneracy between $i$ and
$V_{\rm rot}$, so we kept this parameter fixed for all the rings to its optical value as derived from
Koopmann \etal 2001. 4) The improved values define a new systemic velocity $V_{\rm sys}$ and set of sky coordinates $x_{0},y_{0}$.  Repeat 2) and 3) with these parameters until convergence is obtained.

On average, we have 6 degrees of freedom in each annuli. 
We are aware that points close to the minor axis carry less information about the
rotational velocity, so we used a weighting function of $|\cos(\theta)|$, in order to give more weight
to points close to the major axis.

Stellar rotation curves were obtained for all the galaxies in the sample.
Kinematic parameters derived from the stellar velocities are
presented in Table $\ref{table4}$. Stellar rotation curves are not corrected by asymmetric drift,
since this is beyond of the scope of this paper. This correction requires the solution of the
collisionless Boltzmann equation. Stellar circular velocities for these galaxies are presented
in Cortes \etal (2008). The case of NGC 4064 was complicated since
this galaxy has a strong central bar, and a model based on a pure circular
disk is not fully justified, but this approach is useful for separating the
circular streaming component from the radial streaming component.

We obtained ionized gas rotation curves for six galaxies in the sample.
We were unable to fit a pure circular tilted ring model to the ionized gas
velocity fields for some galaxies due to scarcity of emission (e.g. NGC 4424,
NGC 4606 or NGC 4694), or large non-circular motions (e.g. NGC 4064, NGC 4569,
and NGC 4457). 
However, systemic ionized gas velocities were
estimated for all the galaxies, either by taking the mean of the
averaged LOS velocities at the same
radius on opposite sides along the major axis, or by considering
the LOS velocity at the position of the peak of the stellar continuum map.
The kinematic parameters derived from ionized gas velocities are presented
in Table $\ref{table5}$.

In some of our galaxies, the optical and/or kinematic position angles vary with radius, or there are misalignments between the optical and/or kinematic position angles. We discuss these variations in section 5.1.


\subsection{Comparison with the Literature}

Eight galaxies in our sample have published stellar kinematic data from other groups.
In general the results are consistent, given the difference in spatial resolution
between our DensePak data (3.5") and that for the other groups (0.6"--1.2").

Four galaxies have published long-slit stellar kinematics: NGC 4429
(Simien \& Prugniel 1997), NGC 4450 and NGC 4569 (Fillmore \etal 1986), and
NGC 4698 (H\'eraudeau \etal 1999, Bertola \etal 1999).
Comparison of our data with those from these groups for
the LOS velocity and
velocity dispersion along the major axis is shown in Fig. $\ref{compplot}$.
In NGC 4429, our stellar velocity data show a good correspondence with
the Simien \& Prugniel data, although our velocity dispersions are slightly higher, due to our lower spatial resolution. 
In NGC 4450, there are some disagreements between our data
and the Fillmore \etal data which do not seem attributable to differences in spatial resolution. Since they didn't publish any information about their errors,
we don't know if the differences are significant.
In the case of NGC 4569, the Fillmore \etal data exhibit a peak in the
stellar velocity at 3" that we don't resolve due to our lower
spatial resolution. Both data sets seems consistent in their velocity dispersion
measurements. 
In NGC 4698, our data seem consistent 
with the H\'eraudeau \etal data, within the errors.

Long-slit minor axis stellar kinematics 
published by Coccato \etal (2005) for NGC 4064 and NGC 4424 
seem consistent with our stellar kinematic measurements. 
2-D stellar kinematics for NGC 4293 and NGC 4698 have been obtained by the SAURON project
(Falc\'on--Barroso \etal 2006). Our stellar velocity and velocity dispersion fields are
similar to theirs, with differences inherent to the different spatial
resolutions between SAURON and DensePak (0.94" for SAURON and 3.5" for DensePak). 
Mazzalay \etal (2014) presented a 3"$\times$3" Br$\gamma$ stellar velocity field with SINFONI, with a
high spatial resolution ($\sim$ 0.25"). This map exhibits a higher amplitude in velocity within the inner
3" and a lower velocity dispersion, which is expected due to the huge difference in spatial resolution
between the two maps. 

Long-slit ionized gas kinematics (H$\alpha$) has been published by Rubin \etal (1999). In Fig.
$\ref{compvgas}$, we present a comparison between Rubin's H$\alpha$ long-slit kinematics and
our ionized gas LOS velocities (H$\beta$, and [\ion{O}{3}]$\lambda$5007) along the major axis for nine sample galaxies. Our ionized gas velocities shows a good correspondence with Rubin's data,
although in NGC 4651 our ionized  gas velocities displays an apparent lower rotation velocity. This
could be due to the different spatial resolution between long-slit spatial scale (2") and DensePak (3.5"), which will smear out DensePak velocities. We also have some systematic differences
between [\ion{O}{3}] velocities and H$\alpha$ velocities for NGC 4694, and NGC 4698. We don't know
the cause of this difference. 2-D ionized gas velocity fields have been obtained by Chemin \etal (2006) in H$\alpha$ for NGC 4351, NGC 4450, NGC 4569, and NGC 4580. H$\beta$ velocity fields look similar to his H$\alpha$ velocity fields, but with differences inherent to the different spatial
resolution ($\sim$ 1" for Chemin \etal), and signal-to-noise ratio since H$\alpha$ is much more intense
than H$\beta$ emission. Falc\'on--Barroso \etal (2006) also presented H$\beta$ and [\ion{O}{3}]
ionized gas velocity fields for NGC 4293 and NGC 4698. We were not able to detect any gas emission in NGC 4293. In NGC 4698, both [\ion{O}{3}] velocity fields seems consistent, with
differences inherent to the different spatial
resolutions between SAURON and DensePak. 

The works of Chemin \etal (2006), and Falc\'on--Barroso \etal (2006) have determined kinematic P.A. for some of our sample galaxies.
Chemin \etal (2006) presented H$\alpha$ kinematic PAs for NGC 4351, NGC 4450, NGC 4457, NGC 4569, and
NGC 4580. 
For NGC 4351, we derived (see \S 5.1) 
PA$_{\rm kin}$ H$\beta$ = 62$\deg$ $\pm$ 16$\deg$, which is consistent with 
PA$_{\rm kin}$ H$\alpha$ = 73$\deg$ $\pm$ 5$\deg$ from Chemin \etal (2006).
Both gas kinematic PAs are different from the stellar kinematic  PA$_{\rm kin}$ = 53$\deg$ $\pm$ 8$\deg$.
NGC 4351 is a lopsided galaxy which may be experiencing ram pressure stripping,
and this may account for the difference in kinematic PAs between stars and gas.
For NGC 4450, Chemin \etal give a gas (H$\alpha$) kinematic PA = 171$\deg$ $\pm$ 7$\deg$ in agreement with our
stellar kinematic PA. For NGC 4457, and NGC 4580, Chemin \etal kinematic PAs are in agreement with our stellar kinematic
PA. In the case of NGC 4569, our stellar kinematic PA is 32$\deg$ $\pm$ 6$\deg$, which constrasts with 23$\deg$ $\pm$ 4$\deg$ derived
by Chemin \etal. Our stellar velocity field of NGC 4569 is irregular. The SE side of the velocity
field suggests a different PA than the NW side, perhaps due to dust extinction. The SE side suggests
a kinematic PA closer to the photometric PA of the outer galaxy, so closer to the Chemin \etal PA determination.
The stellar kinematic PA derived in the near-infrared for NGC 4569 (Mazzalay \etal 2014) is about 25$\deg$.
This is seems to confirm that our derivation of the stellar kinematic PA is affected by dust extinction.
Falc\'on--Barroso \etal (2006) give the difference between the photometric and kinematic PAs for
NGC 4293 and NGC 4698. Their values are 30$\deg$ for NGC 4293 and 3$\deg$ for NGC 4698, which
are in agreement with our determination of $\Psi _{\rm inter}$ for both galaxies (see Table $\ref{table4}$).

\section{Observed Stellar and Gas Kinematics}

Figure $\ref{stellarmapset1}$ displays maps
of the absorption and emission-line kinematics,
as well as stellar continuum and emission line intensities
for the thirteen galaxies in our sample.
The maps are displayed in order of increasing NGC number with the same scale and
spatial resolution, and they are oriented with the North up and East to the left.
For each galaxy, we display ({\em first row}) the R-band image from the WIYN Telescope,
maps of the distribution and kinematics of
the H$\beta$ and [\ion{O}{3}]$\lambda$5007 emission lines
for all the galaxies where emission was detected (12/13), and ionized gas residual velocity maps where is possible.
The {\em second row} displays the continuum map
(reconstructed by integrating the spectra within
5250 to 5450 {\AA}), the mean stellar velocity field $V$, the stellar residual velocity map, the velocity
dispersion field $\sigma$, and where is possible $h_{\rm 3}$ and $h_{\rm 4}$ moment maps. 
The open circles in both stellar and ionized gas maps represent broken fibers.
The values at these positions were obtained by
bilinear interpolation, but used only for the purpose of displaying, and they were not used
in the subsequent analysis.
The crosses in the maps represent the positions of the peak in the
continuum emission, and the straight line represents the optical P.A of the
galaxy obtained either by us or Koopmann \etal (2001).
Also, in figure  $\ref{slicesvel1}$ we present the stellar
and ionized gas kinematics along the kinematic major and minor axes.

Detailed descriptions of the different galaxies are collected in Appendix A. Here, we
concentrate on an overview of the general trends of the maps and the resulting
velocity profiles.

\subsection{Stellar and Ionized Gas Velocity Fields}

Stellar velocity fields in the galaxy sample exhibit a variety of
interesting patterns.
Regular stellar velocity fields consistent with pure circular motion
are found in four  galaxies: NGC 4429, NGC 4450, NGC 4580, and NGC 4651.
Other galaxies display misalignments between the photometric and
kinematic major axes, which suggest the presence of non-axisymmetric
structures.
Of these, NGC 4064 has S-shape isovelocity contours and other clear
evidence of a strong stellar bar (Cort\'es \etal 2006), and we suspect NGC 4293 also has a
bar.
Pinching of the central isovelocity contours, suggestive of cold circumnuclear
stellar disks,
are found in NGC 4429 and maybe in NGC 4450.
Finally a remarkable twisting in the isovelocity contours is
found in NGC 4698 which corresponds
to a second kinematical component identified previously as an
orthogonally rotating core by Bertola \etal (1999).

There is also variety in the patterns of stellar velocity dispersion.
NGC 4450, NGC 4569, and NGC 4651 exhibit nearly symmetric
velocity dispersion fields with a clear central peak,
whereas NGC 4606, NGC 4694 and possibly NGC 4698 do not exhibit a clear
central peak within the mapped region.
Other galaxies (NGC 4064, NGC 4293, NGC 4424, and NGC 4429) display more asymmetric $\sigma$ patterns.
NGC 4424 and NGC 4606, which have small bulges and are gravitationally disturbed,
have off-nuclear, off-axis velocity dispersion peaks.
Central drops in velocity dispersion, suggestive of cold circumnuclear stellar
disks,
are present in NGC 4429 and NGC 4694.  The evidence of such a disk is
particularly strong in NGC 4429, which shows pinched
isovelocity contours in the same region.

The ionized gas velocity fields also exhibit a variety of patterns,
and in general look more disturbed than the stellar velocity fields.
Non-circular motions are found in NGC 4064, NGC 4351,
NGC 4457, and NGC 4569.
In NGC 4569, the ionized gas velocity field resembles the H$\alpha$ velocity field for NGC 2992 presented
by Veilleux et al (2001), which shows a clear disk component plus blueshifted and redshifted components
near the minor axis of the galaxy, interpreted as an outflow.
We see similar features in our H$\beta$ and [\ion{O}{3}] kinematic maps of NGC~4569, so 
we suggest that is due to rotation plus an outflow component, although our detected
ionized gas emission is sparse. Note that the high resolution H$_{2}$ velocity field (FOV $\sim$ 2" $\times$ 2")
presented by Mazzalay \etal (2014) also shows a blue shifted region toward the East side of the galaxy and it is
interpreted as an outflow, which is consistent with the apparent blueshift in velocity found in the central
position for the ionized gas (Table $\ref{table5}$) with respect to the stars.
An apparent difference of about 25$\deg$ between the ionized gas
and stellar kinematical P.A. is found in NGC 4450, even though both
patterns are largely consistent with rotation.
Evidence of a twisted gas disk is found in the center of NGC 4698,
in the region of the orthogonally rotating bulge and the transition zone between the bulge-dominated
and disk-dominated regions. 
The kinematic major axis of the gas varies as a function of radius, and 
is very different from the kinematic major axis of the stars.

\subsection{Stellar and ionized gas rotation curves, stellar velocity dispersion
and $V/\sigma$ radial profiles}

Stellar and ionized gas rotation curves, and radial profiles of
stellar velocity dispersion and $V/\sigma$ are shown in
Fig. $\ref{rotcurvset1}$. Residual maps between the observations and
pure circular models are displayed in figure $\ref{stellarmapset1}$. 
The velocity dispersion
profiles were calculated by folding the stellar velocity dispersion along
the kinematic major axis (given by PA$_{\rm kin}$ in Table $\ref{table4}$). 
Errors were estimated as the difference between the measured values and 
the mean of the velocity dispersions on the 2 sides of the galaxy. Velocity dispersion
profiles are not corrected by inclination effects, since this requires
a detailed model of the velocity ellipsoid which is beyond the scope of this work.
The $V/\sigma$ radial profile was calculated as the ratio between the
stellar rotation curve (corrected by inclination), and the velocity dispersion profile at the same
galactocentric radius.

Rotation curves, stellar velocity dispersion and $V/\sigma$ profiles for
all sample galaxies are overplotted in Fig. $\ref{samplekin}$.
The stellar rotation curves span a range of amplitudes and a variety of shapes.
There are galaxies with high stellar rotation velocities (e.g., NGC 4651 or NGC 4429)
and those with extremely low stellar rotation velocities (e.g., NGC 4424, NGC 4351, or NGC 4606),
and this difference reflects the difference in mass between the galaxies. 
But there is also variety in the shapes of the stellar rotation curves for a given galaxy mass.
Whereas some galaxies such as NGC 4064, NGC 4580, and NGC 4651 display a
monotonically rising stellar rotation curve over the entire array,
others  such as NGC 4429, NGC 4450 and NGC 4457
exhibit a maximum within the inner 10" indicating high central mass concentrations.

Stellar velocity dispersion profiles display behaviors as
diverse as the stellar rotation curves. Some
galaxies have high amplitudes in their central velocity dispersion, e.g.
NGC 4429 with $\sigma \sim$ 200 $\kms$.
Other such as NGC 4424 and NGC 4694 exhibit very small amplitudes
($\sigma \sim$ 50 $\kms$). These differences are mostly due to galaxy mass.
Galaxies such as
NGC 4569, NGC 4651, and NGC 4450 exhibit centrally peaked $\sigma$
profiles which decrease outwards, but other
galaxies such as NGC 4698, NGC 4694, and NGC 4606 have essentially
flat velocity dispersion profiles.

Figure $\ref{correl}$ shows the
correlation between the absolute $H$ magnitude, derived using apparent $H$ magnitudes (Gavazzi \etal 1999) and our stellar-kinematics based distances (Cort\'es \etal 2008), and the
maximum stellar velocity rotation,
and the central stellar velocity dispersion. As expected
from the Tully-Fisher, and Faber-Jackson relations, the variations in
$V_{max}$ and $\sigma$ among the
sample galaxies are due predominantly to galaxy mass.

Finally, the $V/\sigma$ ratios show that some galaxies (e.g. NGC 4651, NGC 4569,
or NGC 4580) are largely supported by
rotation ($V/\sigma \geq$ 1 at R$\geq$0.10 $R_{25}$, where $R_{25}$ is the radius
where the surface brightness has a
value of 25 mag arcsec$^{-1}$), but others such as NGC 4424, NGC 4429,
and NGC 4698 have $V/\sigma <$ 1 over the
entire array, and are therefore supported by random motions as
far as we measure.

\section{Kinematic Features and Peculiarities}

While $V$ and $\sigma$ are fundamental measures of galaxy support,
there are important kinematic features of galaxies which are not captured by the radial variations of $V$ and $\sigma$.
These include differences between photometric and kinematic axes, changes in kinematic
position angle with radius (Fig. $\ref{paset}$),
differences in stellar and gas kinematics, and kinematically distinct components. For example;
the presence of kinematic misalignments in galaxies are often used to assess
the frequency of triaxiality, by measuring differences between the photometric and
the stellar kinematics major axes, or to determine the occurrence of accretion events,
by measuring the misalignment between the kinematics of the stellar and gaseous
components (e.g. Falc\'on--Barroso \etal 2006). Changes in position angle may indicate a bar
(Cort\'es \etal 2006).


Many things can cause differences in stellar and gas kinematics.
Some of these are normal features of galaxies which do not directly indicate any galaxy interaction,
such as nuclear outflows, bar or spiral arm streaming motions, and a
different balance of support between rotational and random motions.
In this paper we are more interested in those things which indicate some type of interaction,
such as gas lying in a tilted plane with respect to the stars due to accretion events,
or ram pressure disturbing the gas but not the stars.

Kinematically distinct components are of interest since they can indicate a past merger or accretion event.
Four of our sample galaxies displays signatures of kinematically distinct components,
with the most remarkable cases in NGC 4429 and NGC 4698.

\subsection{Stellar kinematics misalignments}

Both photometric and kinematic position angles can vary with radius.
These radial changes provide important information on the structure and history of the galaxy,
and need to be considered when comparing photometric and kinematic position angles.
In our sample, there are several galaxies which exhibit significant radial changes in photometric or kinematic position angles.
For example, in NGC~4293 and NGC~4569 there are photometric position angle changes from the inner to the outer galaxy associated with
apparent warping of the outer disks.  To handle this complexity, we will compare the stellar kinematic position angles
with photometric PAs measured at more than one radius.
In NGC~4064 and NGC~4450, the stellar kinematic PA varies with radius (Fig. $\ref{paset}$), due to the effect of stellar bars whose
relative importance varies with radius. We will discuss these cases in the text.

We define the inner galaxy stellar kinematics position angle PA$_{kin}$ as the mean of the
kinematic position angles $\phi$ (see Section 3.3), within the inner 25$''$, i.e;

\begin{equation}
{\rm PA_{\rm kin}} =\left< \phi \right>.
\end{equation}
To help deal with the fact that in some of our galaxies, the outer galaxy is disturbed or tilted with respect to the
inner galaxy, we compare this inner galaxy stellar kinematics PA (SKPA) with the photometric PA measured at 2 different radii,
an outer radius of 1.0R$_{25}$, ${\rm PA}_{\rm outer}$, and
an intermediate radius of 0.5R$_{25}$,  ${\rm PA}_{\rm inter}$.
For the photometric position angle determinations,
we use photometry from our own optical images (Cort\'es \etal in prep.).


The kinematical misalignments $\Psi_{\rm outer}$ and $\Psi_{\rm inter}$ are defined as follows (Cappellari \etal 2007):

\begin{equation}
\Psi _{\rm outer} = |{\rm PA}_{\rm kin} - {\rm PA}_{\rm outer}|,
\end{equation}

\begin{equation}
\Psi _{\rm inter} = |{\rm PA}_{\rm kin} - {\rm PA}_{\rm inter}|,
\end{equation}
and are given for all the galaxies in our sample in Table $\ref{table4}$.

For 6 sample galaxies, both $\Psi _{\rm outer}$ and $\Psi _{\rm inter}$ $\leq$7$\deg$
and the degree of misalignment is small.
For these galaxies the inner stellar kinematics exhibit rotational motions consistent with
stellar bodies flattened by rotation. This set includes several galaxies which are nonetheless
clearly disturbed, NGC~4424, NGC~4606, and NGC~4694.

The other 7 sample galaxies have PA differences of $\geq$10$\deg$
for $\Psi _{\rm outer}$, $\Psi _{\rm inter}$, or both.
These galaxies can be grouped into
those with bars or suspected bars (NGC~4064, NGC~4450, and NGC~4293),
those with irregular or disturbed stellar velocity fields (NGC~4351 and NGC~4569), and
those with photometric features not aligned with most of the galaxy (NGC~4429 and NGC~4651).

In both NGC 4064 and NGC 4450, the stellar kinematic PA, $\phi$ varies with radius (Fig. $\ref{paset}$), which we think is
due to strong bar streaming motions whose magnitude changes with radius.
NGC 4064 exhibits a clear difference between the kinematic P.A
and optical P.A angle of about 50$\deg$ in the inner 10", a difference which is
reduced to 10$\deg$ at the end of the array ($r =$ 22"). This produces
a S-type shape in the isovelocity contours.
This galaxy exhibits photometric and kinematical evidence of a bar (Cort\'es \etal 2006) which skews the isovelocity
contours parallel to the major axis of the bar (Athanassoula 1992; Vauterin \& Dejonghe 1997).
Thus, it seems clear that the presence of the bar is the cause of the big
differences between the kinematical and optical P.A.s in NGC 4064.

NGC~4450 seems similar but less extreme, probably because it has a larger bulge.
While $\Psi _{\rm outer}$ and  $\Psi _{\rm inter}$ differ significantly from zero, this is because
the stellar kinematic position angle varies with radius, and
$\Psi$ values are calculated from the mean inner stellar kinematic PA.
In the central 10$''$, it agrees well with the photometric PA of 175$\deg$ in the outer galaxy.
But at $r \sim$15-25$''$, the stellar kinematic PA is closer to 160$\deg$.
This probably reflects a change from the inner bulge-dominated region to the
bar-dominated region at larger radius.

In NGC~4293, the mean stellar kinematic position angle is very
different from the photometric P.A.s, and the reason is not clear.
The outer stellar disk is tilted with respect to the inner disk, indicating a gravitational interaction,
but the stellar kinematic position angle shows a very large difference of $\Psi _{\rm inter}$=35$\deg$
with respect to the photometric PA at both intermediate and small radii (see Appendix).
There is no clear photometric evidence of a bar,
although we cannot rule out a bar extended largely along the line-of-sight.
Moreover, the stellar kinematic major and minor axes are largely perpendicular, and
the velocity field does not exhibit the characteristic S-type shape that it
is expected in a barred galaxy, so there is also no clear kinematic evidence of a bar.
A bar which is extended largely along the line-of-sight may be the reason for the large PA differences
in NGC~4293, although other explanations are possible.

In NGC~4651, $\Psi _{\rm outer}$=10$\deg$ but $\Psi _{\rm inter}$ is small,
because the outer disk is clearly disturbed, whereas the inner galaxy is much more regular
and appears as a normal spiral galaxy.
The outer galaxy contains a stellar tail and shells,
and a disturbed outer \ion{H}{1} distribution which appears warped (Chung \etal 2009),
probably due to a minor merger.
In NGC~4429, the stellar kinematic PA agrees with the photometric PA in both the inner galaxy ($r\leq$25$''$) and outer galaxy,
but differs by 10$\deg$ with respect to the photometric PA at intermediate radii.
This is near the stellar ring at $r \sim $80$''$, which seems tilted with respect to the rest of the galaxy.

In NGC~4569, $\Psi _{\rm inter}$ =13$\deg$$\pm$6$\deg$, which is moderately large. Whereas $\Psi _{\rm outer}$ is smaller,
this is due to the outer disk being tilted with respect to the inner disk, and it does not explain the apparent
difference between stellar and photometric PAs in the inner galaxy.
The inner stellar velocity field is somewhat irregular, possibly due to dust.
The velocity field on the SE side of the major axis suggests a stellar kinematic PA closer to the photometric PA of the inner galaxy,
whereas the NW side, which appears dustier, shows a greater difference.

In NGC~4351, the stellar velocity field is irregular, and there are large values of $\Psi$.
Both may reflect disturbed central stellar kinematics.
The outer isovelocity contours are curved toward the SW, suggesting disturbed stellar motions.
While the stellar kinematic data on this fainter galaxy are somewhat noisy,
similar features are observed in the gas isovelocity contours, suggesting they may be real.

\subsection{Discrepancies between stellar and ionized gas kinematics}

Ionized gas velocity fields are in general more disturbed than the stellar velocity fields.
The dissipative nature of the gas makes it more susceptible to many effects, including
nuclear outflows and shocks triggered by bars or spiral arms.
It also makes it a good tracer for hydrodynamical mechanisms such as ICM-ISM stripping,
which might be revealed by certain differences between the stellar and gas kinematics.
Moreover, the misalignment between the stellar and gas kinematics can be a signature
of accretion events (e.g. Kannappan \& Fabricant 2001).

Misalignments between stellar and gas kinematic position angles
are found in NGC 4351, NGC 4450, and NGC 4698.
In the case of NGC 4450, the misalignment is 25$\deg$ with respect to the kinematic P.A for [\ion{O}{3}].
This is not observed by Chemin \etal (2006) in H$\alpha$ since the H$\alpha$ emission is weak in the center of NGC~4450,
and the H$\alpha$ velocity field of Chemin is noisy. The center of NGC 4450 has been classified as
a LINER (Ho \etal 2000), and accordingly its H$\beta$ emission is very weak, and its [\ion{O}{3}] emission is moderately strong. 
This misalignment could be due to the nuclear activity of this galaxy, although its [\ion{O}{3}] velocity field
seems consistent with rotation rather than an outflow. Alternatively, the gas might be located in a
tilted plane with respect to the stars, which could be produced by an accretion event or minor merger.

NGC 4698 presents an [\ion{O}{3}] velocity field with a kinematic P.A.
different by 30$\deg$ from the stars. This galaxy has been classified as a Seyfert 2 galaxy
(Georgantopoulos \& Zezas 2003), so its unusual ionized gas kinematics
could be related to nuclear activity, but the [\ion{O}{3}]
velocity field is well ordered and seems largely consistent with rotation.
A more likely explanation is that the gas distribution is non-planar,
and the equilibrium plane of the gas gradually changes from larger radii,
where the outer disk dominates the gravitational potential, to smaller radii,
where the orthogonally rotating bulge dominates the potential.

In NGC 4351, the misalignment is about 10$\deg$, and the galaxy
is apparently lopsided (Rudnick \etal 2000), this could suggest a tidal interaction or minor merger. The galaxy
has a nearby neighbor IC 3258 (separation $\sim$ 80 kpc), with a mass of about half of the NGC 4351 mass,
but the relative velocity with respect to NGC 4351 is about 2700 $\kms$,
making the possibility of a strong tidal interaction between the galaxies unlikely. Moreover, the outer stellar disk seems undisturbed.

Alternatively,  the bending of the outer ionized gas isovelocity contours in NGC~4351
could be a signature of ICM-ISM stripping.
A similar bending is observed in NGC 4457, and both galaxies have gas kinematics and
morphologies, which are plausibly caused by ram pressure (see Appendix).

NGC 4569 displays disturbed ionized gas velocity fields with a velocity gradient along the minor axis.
The galaxy nucleus has been classified as a LINER (Keel 1996), and radio maps of NGC 4569 strongly suggest a
nuclear outflow (Chy\.{z}y et al. 2006). H$\alpha$+ [\ion{N}{2}] imaging reveals a C-shaped filament resembling the edges of a cone near the minor axis in the SE, also suggestive of a nuclear outflow (Kenney et al. in prep).



\subsection{Kinematically distinct components}

The existence of kinematically distinct components such as
decoupled cores, cold stellar disks or counter-rotating
components in galaxies could be used as good tracers of environmental
interaction such  mergers or gas accretion.
With the purpose of investigating such
possibilities, we analyzed the line-of-sight velocity distributions
(LOSVD) by obtaining the cross--correlation
function along the major and minor axes of the galaxies using
the IRAF task FXCOR. This task has the advantage
of visualizing the LOSVD independent of the employed parameterization,
which is not possible
with pPXF since we were restricted to obtain $V_{LOS}$ and
$\sigma$ due to signal-to-noise limitations.

We estimate that with our data, we should be able to detect secondary
kinematic components as weak as 20\% of the primary
component in those galaxies with higher signal-to-noise ratio,
and with a velocity difference at least 60 $\kms$ ($\sim 2 \Delta v$,
with $\Delta v$ the
velocity resolution).
No signatures of counter--rotation, like the double peaked LOSVD as in NGC 4550 
(Rubin \etal 1992; Rix \etal 1992) have been found, which could be an indication that galaxies like
NGC 4550 are rare, and are created only under very special conditions.

Signatures of secondary kinematic components other than counter--rotation,
were found in NGC 4429, NGC 4450, NGC 4651, and maybe in NGC 4457.
For NGC 4698, although signatures of an orthogonally rotating core
are obvious from the stellar velocity field, we were
unable to detect any apparent deviation or distinct kinematic
components from the LOSVD profiles.
NGC 4429, NGC 4450, and perhaps NGC 4457 exhibit tails in their LOSVDs
towards the systemic velocity in their inner 10". In the case of NGC 4429 and NGC 4450 these tails have an amplitude
of about 20\% the amplitude of the main component. These tails are evidence of a
two-component structure LOSVD (Rix \& White 1992) within their inner 10", with a hot slow-rotating component
due to bulge and rapidly rotating component. 
NGC 4651 also has tails extending towards the systemic velocity with an amplitude of 30\% the amplitude of the  main component, but
they are visible from $r =$ 10" to 20", rather than in the center as the
other galaxies. This reveals presence of second hot slow-rotating stellar component toward
the outer parts of the galaxy, that
overlaps the rapidly-rotating main stellar component.

In spite of the low signal-to-noise ratio of the spectra (SNR$<$ 60), we obtained
higher moments ($h_{3}$ and $h_{4}$) of the LOSVD
with pPXF, with the objective of investigating the nature of these
asymmetries in the LOSVD. In bulge+disk systems
the shape of the LOSVD can be complicated (Scorza \& Bender 1995),
and even normal galaxies such as NGC 3898, NGC 7782, and NGC 980
(Vega-Beltr\'an \etal 2001)
have non-zero $h_{3}$ and $h_{4}$ moments.
In our case, four moments maps for
NGC 4429, NGC 4450, and NGC 4651 are shown in Fig. $\ref{stellarmapset1}$.
The most remarkable case is
NGC 4429, which shows peaks at 10" in the $V_{LOS}$, a central dip
in $\sigma$, signatures of an anti--correlation
between $h_{3}$ and $V_{LOS}$ in the inner 10", and a positive $h_{4}$
in the inner 10". All these features, together with two-component structure LOSVD, 
are signatures of a cold stellar disk within the inner 10". 
Even the low $V/\sigma$ ratio seems consistent with this scenario, since the bulge dominates the mass
inside the central region. 
We calculate an upper limit on the circumnuclear disk mass as the dynamical mass inside $r=10$",
where the rotation velocity peaks. This disk has a mass of less than 1/4 the mass of 
the galaxy inside $r=25$", where the kinematics and therefore the interior mass are dominated by the bulge.

Similar features are found in NGC 4450, and NGC 4457, although
much weaker. The case of NGC 4651 exhibits an anti--correlation
between $h_{3}$ and $V_{los}$ toward the outer parts of the array
that could be due to a slower rotating component. This is consistent with the presence
of tail toward the systemic velocity in LOSVD. NGC 4698 exhibits very low $h_{3}$ and $h_{4}$ values, even considering
that has an orthogonally rotating core. This is evidence of a relaxed system.
Other galaxies of the sample do not show any apparent deviation from gaussianity or
tails in the LOSVD within the errors.

\section{Kinematical Support and Anisotropy}

\subsection{Velocity dispersion profiles}

The stellar velocity dispersion profiles in stellar disks typically have
exponential profiles with scale lengths about twice the scale length of the light from the disks (Bottema 1993). 
In our case, fitting an exponential profile could be misleading
since our data are restricted to a region where the bulge is important. Instead, we fit a linear profile to our
velocity dispersion profiles $\sigma(R)$ (Table $\ref{table6}$; fig. $\ref{rotcurvset1}$) defined as
\begin{equation}
\sigma(R) = \sigma_{0} + \alpha R,
\end{equation}
where $\sigma_{0}$ is the central velocity dispersion and $\alpha$ is the slope of the velocity dispersion profile.
We found that our galaxies display profiles with slopes ranging from roughly
-36 $\kms$ kpc$^{-1}$ to 6 $\kms$ kpc$^{-1}$.
The galaxies; NGC 4293, 4450, 4569, and 4651 have
$\sigma$ profiles with
slopes steeper than -16 $\kms$ kpc$^{-1}$, but others have very small slopes, with essentially flat profiles. These galaxies are
NGC 4064, NGC 4351, NGC 4424, NGC 4606, NGC 4694 and NGC 4698.
In figure $\ref{dyncorr}$ (top panel), we showed the concentration parameter C$_{\rm 30}$ (related to the 
bulge-disk ratio) with $\alpha$. We don't see a clear correlation, with galaxies having steep or flat velocity
dispersion profiles independent of how conspicuous their bulge are. 

In the group of low $\alpha$ galaxies, NGC 4064, NGC 4424, NGC 4606, and NGC 4694 belong to the group of galaxies
with ``truncated/compact'' profiles H$\alpha$ (Koopmann \& Kenney 2004; section 2). 
NGC 4351 has been classified as ``truncated/normal'' but the low amplitude and flat dispersion profile could be due to the fact
the signal-to-noise ratio of the spectra in this galaxy is not enough to get reliable velocity dispersion measurements. Finally, NGC 4698 is a big
anemic Sa galaxy, which has a clear secondary kinematic component. These
objects are the ones that also have the smallest $V/\sigma$ values within
the central 20" within the sample (Fig. $\ref{samplekin}$). What is the
cause of these flat stellar velocity dispersion profiles?

One possible cause could be that we are measuring $\sigma$, even for the outermost observed position, within the bulge-dominated region.
We checked this, using the available near-infrared photometry and light profile decomposition for
NGC 4606, NGC 4694, and NGC 4698 (Gavazzi \etal 1996), and our own R-band light profile
decomposition for NGC 4064 and NGC 4424 (Cort\'es \etal 2006).
We found that for NGC 4064, NGC 4424 and NGC 4606, the bulge is much smaller than the size of the array
(12", 5" and 9" respectively), so the flat $\sigma$ extends farther than the bulge dominated region. These are
galaxies with less conspicuous bulges (low C$_{\rm 30}$, see fig. $\ref{dyncorr}$) and flat $\sigma$ profiles. In the cases of
NGC 4694 and NGC 4698 the extension of the bulge is 16" and 18" respectively, so it could be that our measurements are still
in the bulge dominated region in these galaxies. In fact, these galaxies have C$_{\rm 30} >$ 0.5, so their bulge is conspicuous.

Another possible explanation for the 
flat stellar velocity dispersion profiles and low $V/\sigma$ values is 
some kind of gravitational interaction.
Mergers and tidal interactions tend to heat galaxy disks and increase the
velocity dispersion (e.g. Bendo \& Barnes 2000). 
Mergers simulations
(Bendo \& Barnes 2000) show that direct encounters
of disk galaxies, with mass ratios of 3:1 can produce
remnants with flat velocity dispersion profiles, slowly rising stellar
rotation curves, and $V/\sigma < 1$ in the inner 20\% of the optical radii.
NGC 4424 and NGC 4698 have remarkably low $V/\sigma$ ratios at the radii
where the disk starts to be important (0.3 and 0.5 respectively). In the case
of NGC 4698, the existence of an orthogonally rotating core clearly
indicated that this galaxy is the result of a merger. The optical morphology
of NGC 4424 has features that suggest an intermediate-mass ratio merger
(Kenney \etal 1996, Cort\'es \etal 2006). The fact that
$V/\sigma$ is remarkably low and flat is consistent with the
merger scenario. With its small bulge, disk-like morphology
and elliptical-like kinematics, NGC 4424 is likely an
intermediate mass-ratio merger (Bournaud \etal 2004; Cort\'es \etal 2006).

NGC 4606 and NGC 4694
exhibit non-elliptical isophotes (Cort\'es \etal in prep.), slowly rising
rotation curves, and low $V/\sigma$ ratios, which resemble the kinematics
found in the Bendo \& Barnes prolate 3:1 merger remnants. 
On the other hand,
NGC 4606 has a close companion NGC 4607 ($d \sim$ 17 kpc, and
$\Delta v =$ 593 $\kms$), which could disturb
this galaxy and produce a flat $\sigma$ profile. The two galaxies have similar
luminosities and masses.
NGC 4607 looks less disturbed than NGC 4606, which might be inconsistent with a strong tidal interaction
between these two galaxies, although if the interaction is prograde for one galaxy (e.g. NGC 4606) and retrograde for the other
(e.g. NGC 4607), it is possible to have one companion significantly more disturbed than the other.
In any case, NGC 4606 has clearly experienced a recent gravitational disturbance.

Finally, the case of NGC 4694 is less compelling
since the bulge is about the size of DensePak array, and
the $V/\sigma$ ratio becomes $\sim$ 1 at end of the bulge dominated region. This
galaxy has a disturbed HI tail connecting with a close dwarf companion
VCC 2062, indicating a clear ongoing gravitational interaction
(van Driel \& van Woerden 1989; Chung \etal 2007; Chung \etal 2009). The velocity dispersion
actually increases slightly with increasing radius, a trend not observed
in any other galaxy. We speculate that this could be the
result of recent star formation in the center of galaxy, which creates a
low dispersion circumnuclear disk.

\subsection{Kinematical Support}

In order to understand the kinematical support of sample galaxies, we made use of the $\lambda_{R}$ parameter (Emsellem \etal 2007), which
is a practical way to quantify the global velocity structure of galaxies using the two-dimensional spatial information provided by the
integral-field units, and for quantifying the specific angular momentum. This dimensionless parameter can be measured via two-dimensional
spectroscopy (Emsellem \etal 2007) as
\begin{equation}
\lambda_{R} = \frac{\sum_{i=1}^{N} F_{i}R_{i}|V_{i}|}{\sum_{i=1}^{N}F_{i}R_{i}\sqrt{V_{i}^{2} + \sigma_{i}^{2}}},
\end{equation}
where $N$ is the number of fibers in the DensePak array, $F_{i}$ is the flux inside the $i$th DensePak fiber, $R_{i}$ its distance to the center,
and $V_{i}$ and $\sigma_{i}$ the corresponding mean stellar velocity and velocity dispersion. The parameter $\lambda_{R}$ is close to unity when
the mean rotation ($V$) dominates. On the contrary, $\lambda_{R}$ goes to zero if the mean velocity dispersion ($\sigma$) dominates. 

The values of $\lambda_{R}$ for sample galaxies are displayed
in table $\ref{table7}$. Following the kinematical classification introduced by Emsellem \etal (2007) and refined by the ATLAS$^{\rm 3D}$ project (Emsellem \etal 2011), all our galaxies can be classified as
{\em fast rotators} with their location in the $\lambda_{\rm R}$ -- $\epsilon$ diagram (fig. $\ref{mergers}$, fig. $\ref{mergersbois}$),
above the so-called ``green line'' (fig. $\ref{mergersbois}$), defined as $\lambda_{\rm R} =$ 0.31$\sqrt{\epsilon}$
(Emsellem \etal 2011), as it is expected for disk galaxies. In fact, all our galaxies show nearly regular and symmetric velocity fields
consistent with being fast rotators. NGC 4698 has the smallest $\lambda_{R}$ = 0.26, and the
highest corresponds to NGC 4651 with $\lambda_{R}$ = 0.71. Also, most of our sample galaxies are located below the
location expected for an isotropic rotator (Fig. $\ref{mergersbois}$, black line), so it is clear that they cannot be
considered simple isotropic oblate rotators as expected.

In figure $\ref{dyncorr}$ (middle panel), we displays $\lambda_{R}$ versus the concentration
parameter $C_{30}$, which is a measure of the bulge-to-disk ratio. We see that galaxies
with low $\lambda_{R}$ tend to have high $C_{30}$ value, although the sample is too small for
deriving further conclusions.
In the case of the relationship between $\lambda_{R}$ and the
slope of the velocity dispersion profile $\alpha$ (fig. $\ref{dyncorr}$; bottom panel),
we see that galaxies classified as
[T/A], [T/N], and [N] can span a wide range of $\alpha$ independent of the value of $\lambda_{R}$.
In the case of the [T/C] galaxies, they have
flat velocity dispersion profiles in spite of their amount of rotational support. The possibility of this being caused by
3:1 or higher mass-ratio mergers is explored in the following section.

\section{Comparison with Simulations}

N-body (e.g. Bendo \& Barnes 2000; Jesseit \etal
2007) and hydrodynamic simulations (e.g. Kronberger \etal 2007; Kronberger \etal 2008)
have started to produce 2-D kinematics that can be compared with the observed velocity
and velocity dispersion fields of galaxies. In order to search for clues about what
physical processes have acted on these galaxies, and which ones are driving galaxy
evolution, we compare our observed kinematics with the results of simulations
of gravitational interactions (i.e., mergers and tidal encounters) and ICM--ISM stripping. 

\subsection{Gravitational Interactions}

Gravitational interactions include mergers and non-merging gravitational encounters.
In clusters, most of the non-merging gravitational encounters will be high-velocity encounters,
sometimes called harassment. Slow encounters can also be important, since gravitationally bound galaxies may interact and later become unbound through a gravitational interaction with another object.
It can be hard to distinguish minor mergers, low-velocity encounters and high-velocity encounters 
without detailed information, since many of their impacts are similar.
While all have been simulated,
we are not aware of any suite of simulations that have included the variety of gravitational interactions and attempted to distinguish between them. 
In the following we compare the dynamical heating and kinematic properties 
of our Virgo sample mostly with merger simulations,
since the results of such simulations exist in a form that can be easily compared with our data.
We discuss simulation results from non-merging gravitational encounters where possible.

\subsubsection{Dynamical Heating of Sample Galaxies}

The results of the SAURON and ATLAS$^{\rm 3D}$ projects show that the 
possible merger origin of galaxies can be assessed 
by comparing the location of a galaxy sample (figure $\ref{mergers}$)
in the $\lambda_{R}$--$\epsilon$ plane measured within the effective radius,
with the location for merger remnants given by simulations (e.g. Jesseit \etal 2009; Bois \etal 2011). 
Our observations reach a radius of 25", which in most of the cases
is less than the half-light radius (table $\ref{table7}$). Thus, in order to carry out this comparison
it is necessary to estimate $\lambda_R$, and the luminosity-weighted ellipticity $\epsilon$ at half-light radius (equivalent to the effective radius for spiral galaxies).
To do this estimation, we made use of our R-band light profiles, and two-integral dynamical models (Cort\'es \etal 2008)
as follows;

\begin{itemize}
\item We estimated the half-light radius $R_{1/2}$ from our R-band light profiles.
\item We derived $\lambda_R$ at $R=$ 25" and at $R=R_{1/2}$ form our Jeans isotropic models as it is described by equation 4.
\item We estimated $k = \lambda_{\rm R}({\rm obs})/\lambda_{\rm R}({\rm iso})$, 
where $\lambda_{\rm R}({\rm obs})$ is the observed $\lambda_{\rm R}$ at $R=$ 25", and $\lambda_{\rm R}({\rm iso})$ is $\lambda_{\rm R}$
derived from our Jeans isotropic model at $R=$25". This ratio is analog to the so-called Satoh's parameter (Satoh 1980),
and it is an estimation of the tangential anisotropy. Although, the interpretation of this ratio in real galaxies is not
so simple, since this can be tie to the galaxy's inclination and the vertical anisotropy ($\delta$) (e.g. Capellari
\etal 2007).
\item $\lambda_{\rm R}$ at $R_{\rm 1/2}$ was estimated as $\lambda_{\rm R_{\rm 1/2}}= k \lambda_{\rm R_{\rm 1/2}}({\rm iso})$, where $\lambda_{\rm R_{\rm 1/2}}({\rm iso})$ is $\lambda_{\rm R}$ at $R_{\rm 1/2}$ derived from our Jeans isotropic models.
We recall that this is just an estimation of $\lambda_{\rm R}$ at $R_{\rm 1/2}$, since this depends that $k$ remains
constant within the galaxy, which may not be the case.
\item Finally, we calculated the ellipticity $\epsilon$ at $R_{\rm 1/2}$ from our surface brightness models used
to derived our Jeans isotropic models, as it was done in Capellari \etal 2007. That is by calculating the
luminosity-weighted ellipticity within the inner $R=R_{\rm 1/2}$. 
\end{itemize}

Correction to $R_{\rm 1/2}$ displaces the location of the galaxies mostly toward
higher ellipticities and higher $\lambda_{\rm R}$.
Values for both the measured and corrected values are given in Table $\ref{table7}$.  
In Figures 13 and 14 the measured and corrected values are connected by lines, with the symbols located at the corrected values.

We compare the location of our galaxies in the $\lambda_{R}$--$\epsilon$ plane (figure $\ref{mergers}$)
with the location of simulated merger remnants  from Jesseit \etal (2009) and Bois \etal (2011), in figures 13 and 14, respectively.
The simulations of Jesseit \etal (2009) are disk-disk mergers with mass ratios between 1:1 and 3:1,
for both gasless (``dry") mergers and those with gas and star formation (``wet").
The simulations of Bois \etal (2011) are ``wet" spiral-spiral mergers with mass ratios of either 3:1 or 6:1.
The ``progenitor" galaxies of Bois \etal (2011) are dynamically hotter than those in  Jesseit \etal (2009) 
since they already include bulges, with bulge-to-disk ratios corresponding to Sb and Sc galaxies.
Thus, the Bois \etal (2011) progenitors represent galaxies that have likely already experienced gravitational disturbances.
Simulations show that wet mergers produce more rounded remnants than dry mergers,
especially in the low rotator region. 
All of our sample galaxies have gas, so the comparisons with the wet mergers are most appropriate.

Our sample galaxies occupy 3 main regions in the $\lambda_{R}$--$\epsilon$ plane, thus we
refer to 3 groups: dynamically cold galaxies, dynamically lukewarm galaxies, and dynamically warm galaxies.

1. Dynamically cold galaxies.
Two galaxies, NGC 4580 and NGC 4651, are
consistent with the Jesseit \etal (2009) progenitor line, and well above the Bois \etal (2011) progenitors.
In the case of NGC 4580, the stellar component of the galaxy looks undisturbed, 
so it probably has not been affected by any significant gravitational interaction.
NGC 4651 displays a tidal tail and warped \ion{H}{1} outer disk (Chung \etal 2009),
suggesting an ongoing minor merger, but either its mass ratio must be very high or the merger is in an early phase, leading to a kinematically undisturbed inner disk.    

2. Dynamically lukewarm galaxies. 
Most of our sample galaxies occupy a central band in the $\lambda_{R}$--$\epsilon$ plane,
extending from ($\epsilon$, $\lambda_{\rm R}$)$\simeq$(0.1,0.25) to (0.7,0.55).
These lie well below the Jesseit \etal (2009) progenitor line, and a bit below the Bois \etal (2011) mean progenitor line.
Some are in regions consistent with 3:1 or 6:1 merger remnants (NGC 4694, NGC 4450, NGC 4457, NGC~4429 and NGC 4698)
although some with high ellipticitites are beyond any of the merger regions (NGC~4293, NGC~4064, NGC~4569, NGC~4606).
Figure 13 shows that all of the high ellipticity galaxies have weak bulges. At least some of these have bars.
All of the galaxies in this group have likely experienced at least modest gravitational disturbances.
While some could be 3:1 to 6:1 mergers, many could be minor mergers ($\geq$10:1) or tidal encounters, especially those with high ellipticity.

3. Dynamically warm galaxy. 
One galaxy in our sample, NGC~4424, is dynamically much warmer than any of the progenitors from either 
Jesseit \etal (2009) or Bois \etal (2011), and in a region consistent with a 3:1-6:1 merger remnant.
Its rotational support is much less than any other sample galaxy, 
and it has been previously called a ``low rotator" (Rubin \etal 1999),
although it has enough rotation to qualify as a ``fast rotator" as defined by Bois \etal (2011).
Given its peculiar morphology and kinematics, NGC~4424 has been previously proposed to be a merger remnant
(Kenney \etal 1996; Cort\'es \etal 2006).

It is of interest to compare the dynamical state of the galaxies with their star formation classes. The only [N] galaxy in the sample is the dynamically coldest one. One of the three [T/N] galaxies, NGC~4580, is dynamically cold, providing further evidence that its truncated gas disk is due to ram pressure stripping rather than a gravitational interaction (Koopmann \& Kenney 2004).
The other two [T/N] galaxies, NGC~4457 and NGC~4569 are in the dynamically lukewarm band, and may have experienced gravitational interactions in addition to unrelated ram pressure stripping events. For the [N] and [T/N] galaxies, 
the star formation classes are related to their dynamical state in the expected way.

Galaxies classified as [T/A] are all in the dynamically lukewarm band. All have bulges, so
the dynamically warm components may be bulges. Some of these may have had non-merging
gravitational encounters, but others likely are the result of higher mass ratio mergers
(mass ratios $\geq$ 1/6). This is consistent with the existence of kinematically distinct
components in NGC 4450 and
NGC 4429 and an orthogonally rotating core in NGC 4698. It has been unclear what has
happened to these galaxies to give them [T/A] star formation distributions. Since they are
all consistent with having gravitational disturbances, it is possible that the
gravitational disturbances are responsible for the [T/A] distributions.

Galaxies classified as [T/C] are all in the dynamically lukewarm band, except for NGC~4424
which is dynamically hotter. [T/C] galaxies all have relatively high ellipticities.
They all have small bulges, at least some have bars, plus they happen to be viewed at a high
inclination angle.  It is notable that the dynamically warm or lukewarm component of these
galaxies are not bulges, but disks. The high ellipticities are inconsistent with lower mass
ratio mergers (mass ratios $\leq$1/6), so these galaxies have experienced either higher mass
ratio mergers (mass ratios $\geq$1/6) or non-merging gravitational encounters. In terms of
dynamical heating the [T/C] galaxies are similar to the [T/A] galaxies but they have lower
masses.  The [T/A] galaxies have an average $H$-band absolute magnitude $\sim$ -23.5 whereas
the [T/C] galaxies have H$\sim$ -22.

\subsubsection{Kinematic Peculiarities}

N-body simulations (Jesseit \etal 2007) show that merger remnants display a high diversity in
their 2D maps of various moments of the LOSVD. They display features such as 
low rotation, kinematical twists and misalignments, counter-rotating cores, $v$--$h_{3}$ correlations or anti-correlations, polar rings, low-$\sigma$ rings, and central $\sigma$ drops. 
Mergers are not the only gravitational interaction to cause dynamical heating and kinematics peculiarities.
Recent chemodynamical numerical simulations (Bekki \& Couch 2011) consider that slow tidal encounters
of galaxies and groups tides could be responsible of the transformation of spirals into S0s. 
These simulations show that resulting galaxies display lower rotational
velocities than their spiral progenitors ($V/\sigma \sim$ 1), flat radial $\sigma$ profiles, 
or central dips in $\sigma$.
High velocity encounters may also cause similar features.
Here, we will refer to the sample galaxies that exhibit these telltale features.\\
\\
{\bf Low rotation and flat $\sigma$ profiles.} Flat $\sigma$ profiles and lower rotation amplitude ($V/\sigma \sim$ 1) 
can be caused by either minor mergers (Bournaud \etal 2004; Jesseit \etal 2007)) or tidal encounters (Bekki \& Couch 2011).
Here we have only one galaxy, NGC 4424, which has a rotation velocity below
30 $\kms$. Low rotation can be created in a 1:1 merger, but this is not likely for NGC 4424
since this galaxy exhibits an exponential light profile, which seems to contradict a 1:1 merger
scenario. This kind of object with low rotation and an exponential light profile can be
the result of higher mass ratio mergers (1:10 - 1:4; Bournaud \etal 2004).
Flat $\sigma$ profiles and lower rotation amplitude are also observed in 
NGC 4064, NGC 4606, NGC 4694 and NGC 4698. \\
\\
{\bf Kinematical misalignment and twist.} In our galaxy sample we don't find any galaxy with
big kinematical misalignments ($\Psi >$ 30 $\deg$), which are produced in 1:1 mergers. 
We do have one sample galaxy NGC 4293 with an intermediate amount of misalignment, $\Psi \leq$ 20-30 $\deg$.
In the case of 3:1 mergers, simulations (Jesseit \etal 2007) show that $\Psi \leq$ 20 $\deg$. 
These are in better agreement with most of our galaxies.
In the case of NGC 4651, $\Psi_{outer}=$ 10$\deg$ $\pm$ 4$\deg$ (table $\ref{table4}$), which is consistent with 
a minor merger. The most remarkable case of kinematical misalignment is NGC 4698 with its well known orthogonally rotating core.
This galaxy could be considered similar to polar rings galaxies, which can be formed by mergers with a dissipative component.\\
\\
{\bf Counter-rotating core.} We don't find any sample galaxy with this kind of feature,
although the orthogonally rotating core in NGC~4698 clearly implicates an old merger, and 
a misaligned gas core in NGC~4424 (Cortes \etal 2006) strongly suggests a merger.\\
\\
{\bf $v$--$h_{3}$ correlation or anti-correlation.} Simulations show that a positive correlation between
these quantities, is created in the center of collisionless 3:1 remnants (Jesseit \etal 2007). The presence of gas
suppresses the formation of this correlation, and leads instead to an anticorrelation between $v$--$h_{3}$.
We see suggestions of such an anticorrelation in NGC 4429, NGC 4450 and NGC 4651.\\
\\
{\bf Double $\sigma$-peaks.} This feature has been seen in simulations of some equal-mass merger remnants
with gas (Jesseit \etal 2007). In our sample two galaxies display clearly off-axis double peaks,
similar to what is seen in simulations. These are NGC 4424, and maybe NGC 4606. Both galaxies have
truncated/compact H$\alpha$ morphologies, with active star formation in the center. These feature suggest
that these galaxies are indeed the result of mergers, but that such 
features are not unique to equal-mass mergers but can be present in non-equal mass mergers. \\
\\
{\bf central $\sigma$ drops.} In simulations of both mergers (Jesseit \etal 2007) and tidal interactions (Bekki \& Couch 2011) of galaxies containing gas, the center often exhibits low $\sigma_{\rm los}$ due to significant amounts of gas driven to the center
which then forms stars.
We found this in NGC 4429 and NGC 4694. NGC 4429 shows a clear $\sigma$ drop in
the center, as well as other features (see \S 5.3) that suggest a cold circumnuclear stellar disk, which likely formed after most of the bulge component.
NGC 4694 exhibits a small central dip in $\sigma$. The results of two-integral
self consistent dynamical models (see Cort\'es \etal 2008) show a low $M/L$ $\sim$ 1.0 $M_{\odot}/L_{\odot}$, together with its $B$ -- $V$ color ($\sim$ 0.62) suggesting a recent starburst, which could explain the $\sigma$ dip in the center.\\

\subsection{ICM--ISM Stripping}

While many Virgo cluster spiral galaxies show strong evidence for ongoing ram pressure stripping 
(Koopmann \& Kenney 2004; Chung \etal 2007, 2009),
we focused on galaxies with other types of peculiarities in selecting the present sample.
Nonetheless we included a couple of galaxies thought to be good examples of ongoing or past ram pressure stripping,
NGC 4569 and NGC 4580. Two other galaxies in the sample, NGC 4351 and NGC 4457, may be experiencing ram pressure.
We discuss these galaxies below.
In addition, other galaxies in the sample may have experienced ram pressure stripping, in addition to gravitational encounters,
since they are gas-poor, although we don't have evidence in our data for ongoing ram pressure interactions.

Active stripping is easier to identify in highly-inclined galaxies, since a clear spatial separation of gas and stars can be seen
(Kenney \etal 2004; Crowl \etal 2005; Abramson \etal 2011).
In relatively face-on galaxies, a comparison of stellar and gas kinematics is a good way to identify ongoing ram pressure stripping.
Ram pressure acts only on gas, so in a galaxy affected by ram pressure,  the kinematics of the gas is disturbed, but those of the older stars is not.
If however star formation occurs in gas disturbed by ram pressure, the stellar kinematics of the newly-formed stars would be disturbed, so the 
intensity-weighted stellar kinematics (including both young and old stars) could also be disturbed.
Recent (e.g., Kronberger \etal 2008), combined N-body/hydrodynamic simulations have shown that
ICM--ISM stripping has distinctive effects on the gas velocity fields, which become distorted. 
In the case face-on stripping, velocity fields look
disturbed in the outer parts and undisturbed in the inner parts of the disk. As the inclination of the
galaxy becomes more edge-on, the disturbance in the gas velocity field increases and it becomes asymmetric,
with a mismatch between the gas kinematic center and the stellar disk center.

Some of our sample galaxies display features in their ionized gas
velocity fields which may indicate ram pressure.
NGC 4457 displays a stellar velocity field largely consistent with rotation, but
an asymmetric ionized gas velocity field indicating non-circular gas motions.
There is an apparent bending in the gas isovelocity contours on the southern side of the galaxy, and 
gas velocities are 30-40 $\kms$ lower than the stellar velocities. This bending is also visible in the more
detailed and extended map presented by Chemin \etal (2006).
Also, its gas kinematic center doesn't seem
to coincide with the optical center of the stellar disk, although the gas kinematic center
is not well determined. Moreover, it presents a peculiar
H$\alpha$ arm. ICM--ISM stripping simulations (e.g. Schulz \& Struck 2001), show that
this feature can be created in galaxies tilted with respect to the ICM wind.
Overall, the disturbed gas kinematics and anomalous H$\alpha$ arm suggest active ram pressure stripping in NGC~4457.

NGC 4351 is lopsided, as the starlight in the central $\sim$~30"
is significantly offset from the centroid of outer disk isophotes.
Much of the central starlight arise from strong star formation with an irregular pattern.
Otherwise, there are no obvious disturbances in the light distribution of the outer galaxy,
which lacks star formation. The stellar velocity field is generally consistent with a rotation pattern
(Fig. $\ref{stellarmapset1}$), except that all the isovelocity contours curve toward the SW.
While not all of the features which deviate from a pure
rotation pattern are real, the unusual curvature  toward the SW may be real.
A similar pattern is observed with high significance in both ionized gas velocity fields (Fig. $\ref{stellarmapset1}$),
as well as the HI velocity field (Chung \etal 2009).
It remains unclear whether this apparently similar disturbance to the stellar and gas velocity fields is due
to a tidal interaction or ram pressure.
While ram pressure doesn't affect stars, much of the stellar light we are measuring in this region
could be from young stars which formed in gas disturbed by ram pressure.
The HI distribution shows compressed contours to the NE, and extended, tail-like distribution to the SW (Chung \etal 2009).
Similarly, the peak in ionized gas intensities is in the NE.
All these features may be consistent with ongoing ram pressure from the NE.

NGC 4580 has a sharply truncated gas disk within a relatively normal looking stellar disk.
Its high V/$\sigma$ ratio confirms that the stellar disk is undisturbed.
Its ionized gas velocity field looks symmetric and undistorted, indicating that strong ram pressure is not presently acting on the galaxy.
The H$\alpha$ image of the galaxy shows a prominent star forming ring at the gas truncation radius (Koopmann \& Kenney 2004).
This could be a case of ``annealing'' (Schulz \& Struck
2001), where some of the outer disk gas has lost angular momentum and forms a dense gas ring at the truncation radius which 
makes the remaining gas more resistant to stripping.

NGC 4569 may be the clearest case of ongoing ram pressure stripping in our sample, although
our kinematic data don't show anything obviously related to this since the telltale evidence is beyond the central field of view we studied.
It has a sharply truncated gas disk within a relatively normal stellar disk, as well as
an anomalous extraplanar arm of HI gas with star-forming regions which may be falling back into the galaxy.
Vollmer \etal (2004) make the case that NGC4569 is in a post-peak pressure phase, after closest approach.
The kinematic peculiarities we find in the central part of NGC 4569 reflects a nuclear outflow, and is not directly related to the ram pressure stripping.


\section{Peculiar Cluster Galaxies and Environmental Effects}

The kinematic peculiarities of the sample galaxies are difficult to explain by internal mechanisms, and
are presumably due to interactions within the cluster or circumcluster environment.
Can these peculiarities be caused by a
sole mechanism, or do we need more than one mechanism acting independently
for explain them? In order to answer these questions, we focus on the probable
scenarios for the galaxy peculiarities suggested by
the stellar and ionized gas kinematics for all the sample galaxies;\\

In Appendix A we give a detailed discussion on the kinematics of each sample galaxy
and in table $\ref{table8}$ we summarize the kinematic and morphological indicators of interactions for each galaxy,
and give our best guess for the types of interactions that the galaxies have experienced.
All galaxies show evidence for ongoing, recent, or past interactions,
and several galaxies show evidence for multiple interactions.

Two galaxies in our sample, NGC 4351 and NGC 4457, show possible evidence of ongoing ram pressure stripping
based on our kinematic data.
One galaxy proposed to be an example of past ram pressure stripping, NGC 4580 (Koopmann \& Kenney 2004, Crowl \etal 2008)
is confirmed in our kinematic data to be undisturbed and have a kinematically cold stellar disk,
inconsistent with any past strong tidal interaction and perfectly consistent with the past ram pressure stripping scenario.
All but one of our galaxies, NGC 4651, are very gas poor (Chung \etal 2009) and may have experienced gas stripping.
Several of these galaxies show clear or possible evidence for ongoing or past stripping
based on HI mapping (Chung \etal 2009), but not from our data.

Among the most intriguing objects in our sample are those four galaxies with truncated/compact H$\alpha$ morphologies.
All have flat stellar velocity dispersion profiles.
While NGC~4064, NGC~4424, and NGC~4606 all have small bulges, all have much lower angular momentum parameter
$\lambda _{\rm R}$ than NGC~4580, a small-bulge galaxy with similar mass, indicating that their stellar disks have been heated.
The only one of the four with a prominent bulge, NGC~4694,
has a central dip in its velocity dispersion, suggesting a young circumnuclear stellar disk.
These four galaxies have likely experienced gravitational interactions
which have disturbed the stellar disks and driven gas to the central kpc.
NGC 4424, with the strongest kinematic and morphological disturbances,
likely experienced a recent  intermediate-mass ratio merger (Kenney \etal 1996; Cort\'es \etal 2006)
or a close high-velocity collision with another galaxy.
NGC 4694, with a complex HI distribution including a tail connecting to a nearby dwarf,
clearly has experienced a galaxy-galaxy tidal interaction, probably involving accretion or a ``partial merger'', that is that the nearby dwarf got disrupted and part of it merged with NGC 4694.
The detailed type of gravitational interaction is not as as clear in NGC 4606 and NGC 4064,
but their kinematic properties suggest intermediate-mass ratio mergers or close tidal encounters.
These galaxies have probably also experienced ICM-ISM stripping since they are gas-poor, have very truncated
HI and H$\alpha$ disks, and in the case of NGC 4424 an HI tail (Chung \etal 2007; Chung \etal 2009).

Several of the sample galaxies have kinematically distinct cores,
which result from gas infall followed by star formation, and are likely signatures of accretion or mergers.
NGC~4698, with its orthogonally rotating bulge, is clearly an old merger.
NGC~4429, with its misaligned central stellar and dust disk, and the tail in its LOSVD, is likely an old merger due
to the lack of gas observed in this galaxy, which probably was consumed in the subsequent star formation that
formed the kinematically distinct core.
NGC~4450, with the tail in its LOSVD, and the kinematically misalignment between its central gas and stars,
has likely experienced a recent minor merger.
NGC~4694, with its central sigma drop and evidence for a young central stellar disk, plus the clearly disturbed HI,
has likely experienced accretion or a partial merger with the dwarf galaxy it is currently interacting with.
While it doesn't have a kinematically distinct core, the disturbed outer disk of NGC~4651 clearly implicates a recent minor merger,
and the the prominent tail in its stellar LOSVD at r=10-20$''$ suggests that the merger has affected the inner galaxy.
The old mergers could have happened before the galaxies entered the central part of the cluster.
Among the galaxies with evidence for recent minor mergers, NGC~4651 is far behind the main cluster,
and NGC~4694 and NGC~4450 have at 3D distances beyond, but not far from, the virial radius of the cluster.

Non-merging tidal encounters are likely responsible for the misaligned inner and outer disks in 
NGC~4569 and NGC~4293, as well as  their central kinematic and morphological PA offsets.
Fast tidal encounters are expected to be common in clusters, and may be responsible 
for some of the other kinematic peculiarities we have observed. 
Some features can be produced by either mergers or non-merging tidal encounters, and 
their origin can be hard to pinpoint without a fuller picture of the galaxies properties.

The fact that most (10/13) of our sample galaxies have signatures
of some kind of gravitational interaction shows that they play
important roles in driving cluster galaxy evolution. 
At least half of our sample galaxies show evidence for ICM-ISM stripping, 
a fraction which underestimates the general importance of gas stripping in clusters, since
we chose mostly galaxies which exhibited features not consistent with simple ICM-ISM stripping, 
Thus, a significant fraction of cluster galaxies experience both gravitational
interactions and ICM-ISM stripping.

\section{Summary}

In this paper, we have studied the stellar and ionized gas kinematics
of 13 bright peculiar Virgo Cluster galaxies to
investigate the mechanisms responsible of their peculiarities. Our results are
summarized as follows:\\
\\
1. We present 2-Dimensional maps of the stellar velocity
$V$, and stellar velocity dispersion $\sigma$  and the ionized gas velocity (H$\beta$ and/or [\ion{O}{3}])
of the central 30$''$$\times$45$''$ for galaxies in the sample. 
We show maps of the higher-order moments of the LOSVD $h_3$ and $h_4$ for the 3 galaxies with the highest signal-to-noise ratio.
The stellar rotation curves and velocity dispersion radial profiles are determined for 13 galaxies,
and the ionized gas rotation curves are determined for 6 galaxies. 
We show deviations from circular gas and stellar motions by subtracting models of circular motions from the data.
We measure the kinematic position angles for the gas and stars, and compute
differences between the optical and kinematical major axes.
We provide improved position angle and inclination values for some galaxies.\\
\\
2. The stellar velocity fields in the centers of our sample galaxies are largely consistent with a rotational pattern,
except for the strongly barred NGC 4064, and
NGC~4698, which has an orthogonally rotating core.\\
\\
3. NGC 4064 exhibits an S-type pattern in its stellar velocity field, and
big differences between the optical and kinematical P.A within the inner 10$"$, 
due to strong non-circular motions caused by a central bar.
The bar in NGC 4064 is unusually strong. It may be the strongest stellar bar observed to date, in terms
of deviations from circular motions for the stars.
Large misalignments between the
optical and kinematical major axes are also found in the stellar velocity field of NGC 4293,
however in this galaxy the presence of a bar is not clear.\\
\\
4. Nine galaxies exhibit non-circular gas motions and/or
systematic differences between the stellar and ionized gas kinematics.
The causes of these anomalies are varied.
There is a velocity gradient along the minor axis caused by a nuclear outflow in NGC~4569.
There are large differences between the kinematic P.A. of the gas and stars,
in NGC~4698 and NGC~4450, even though both gas and stars have predominantly circular motions.
We propose this is caused by the
gravitational interaction of the gas with the potential of the disk and orthogonally rotating bulge in NGC 4698,
and recent gas accretion or minor merger in NGC~4450.
Two relatively face-on galaxies in our sample, NGC 4351 and NGC 4457, show possible evidence of ongoing ram pressure stripping
based on distinctive curvatures in gas isovelocity contours consistent with ram pressure stripping.\\
\\
5.
Stellar velocity dispersion radial profiles exhibit a range of behaviors.
Some galaxies have sharply decaying profiles, including NGC 4569, NGC 4651, and NGC~4450.
Some have roughly flat profiles, such as NGC 4424, NGC 4606, NGC 4064, and NGC 4580.
Others have more complex profiles, with maximum values outside the center, including
NGC~4698, NGC~4429, and NGC~4694. These galaxies have kinematically distinct cores.
There is not a one-to-one correlation with bulge-to-disk ratio, as
there are galaxies with both large and small bulge-to-disk ratios that have relatively flat profiles.\\
\\
6. The $V/\sigma$ ratios are bigger than unity where
the exponential disk starts to dominate the light in most but not all of the galaxies.
Galaxies such as NGC 4569, NGC 4580, and NGC 4651 are
largely supported by rotation ($V/\sigma \geq 1$ at 0.05 $R_{25}$),
but others such as NGC 4424, NGC 4429, and NGC 4698
have $V/\sigma < 1$ over the entire array, so they are supported
by random motions as far out as we can measure.\\
\\
7. Signatures of kinematically distinct stellar components have been found in several galaxies.
Pinched isovelocity contours and analysis of the higher moments of the LOSVD ($h_{3}$, and $h_{4}$), reveal
cold circumnuclear stellar disks in NGC 4429 and maybe in NGC 4450.
In the case of NGC 4429, this disk has a mass of 1/4 the mass of the bulge, and is misaligned with the outer disk,
suggesting a merger origin.
NGC~4694 has a central $\sigma$ drop and evidence for a young central stellar disk, probably related to an ongoing
gravitational interaction.
The stellar velocity field in NGC 4698 clearly reveals the existence
of an orthogonally decoupled core (Bertola \etal 1999), undoubtedly due to an old merger.
While it doesn't have a kinematically distinct core, the disturbed outer disk of NGC~4651 clearly implicates a recent minor merger,
and the prominent tail in its stellar LOSVD at r=10-20$''$ and an anticorrelation between  $V$--$h_{3}$ 
suggests that the merger has affected the inner galaxy.\\
\\
8. We have found no evidence of stellar counter-rotating components
at least as big as 20\% of the primary component. This
indicates that disk galaxies with massive counter-rotating components are rare,
and are only created under very special conditions.\\
\\
9.
We have computed for sample galaxies the angular momentum parameter $\lambda_{R}$,
which describes the relative amounts of rotational and random stellar motions.
Only 2 of our galaxies are consistent with being strongly dominated by rotation.
Random stellar motions are dynamically important in most of our galaxies,
indicating the importance of minor to intermediate mass ratio mergers and 
gravitational interactions in establishing their dynamical states.\\
\\
10.
Several galaxies with small bulges and truncated/compact H$\alpha$ morphologies
have low ratios of V/$\sigma$, flat radial $\sigma$ profiles, and values of $\lambda_{\rm R}$
that indicate that their stars have been dynamically heated by gravitational interactions.
This may include 3:1 or higher mass-ratio mergers, which aren't as disruptive as 1:1 mergers, 
allowing the galaxies to retain their fast rotator characteristics and disk-like morphology, but puffing up the disks.
While ram pressure stripping may be partly responsible for the truncated gas disks in these truncated/compact 
galaxies, their properties cannot be fully explained by simple ram pressure stripping, and require gravitational interactions.\\
\\
11. Comparison with simulations reveals that most (10/13) of our sample galaxies have signatures
of some kind of gravitational interaction, showing that they play
important roles in driving cluster galaxy evolution. 
At least half of  the galaxies show evidence for ICM-ISM stripping, 
and several galaxies exhibit evidence for both.
This implies that a significant fraction of cluster galaxies experience both gravitational
interactions and ICM-ISM stripping.

\bigskip
\bigskip

The funding for this research has been provided by Fundaci\'on Andes, Chile;
Fondap project grant 15010003, Chile.
This work is part of the research on the Virgo Cluster
and isolated spiral galaxies funded by NSF grant AST-0071251.

\appendix
\section{Description of Individual Galaxies}

In this section, we comment on the velocity and velocity dispersion fields for each galaxy,
as well as some relevant background information from other sources.
We also explain the peculiarities, H$\alpha$ type components and features
within the central $r\sim$25" region mapped by us.\\ 

{\bf NGC 4064:} NGC 4064 has a strong stellar bar which turns into very open spiral arms at larger radii.
Star formation is very strong in the central $\sim$1 kpc, but virtually absent at larger radii.
Disturbed dust lanes extend to $\sim$ 3 kpc  (Cort\'es \etal 2006).
These features suggest some kind of gravitational interaction, plus 
a process that removes the outer disk gas. 

\begin{figure}
\figurenum{2}
\vbox{
\begin{center}
\includegraphics[height=4.5in]{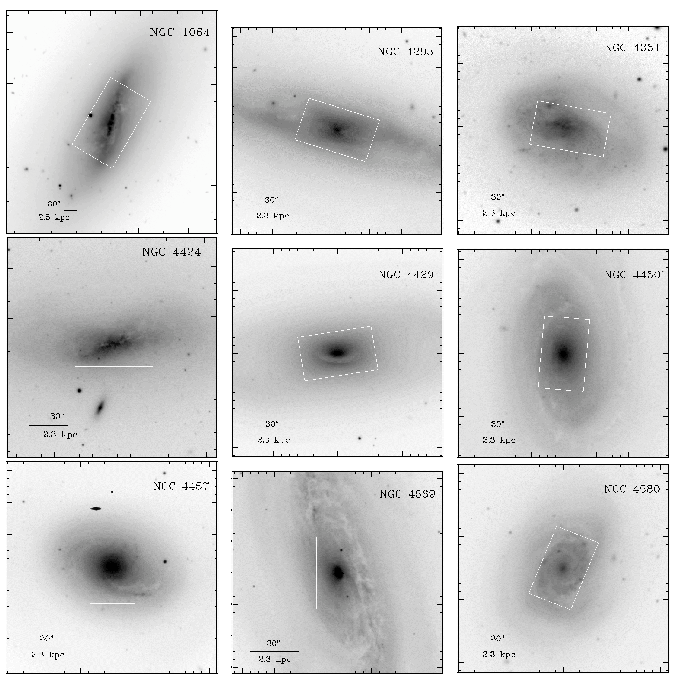}
\end{center}
}
\caption{R-band WIYN mini-mosaic images for some sample galaxies and the DensePak array.
mage sizes is about 2'$\times$2', with the usual orientation North up and East
left. White rectangle represents the DensePak array as it was oriented in the sky.}
\label{galaxydspk1}
\end{figure}

\begin{figure}
\figurenum{3}
\begin{center}
\leavevmode
\includegraphics[width=5.0in]{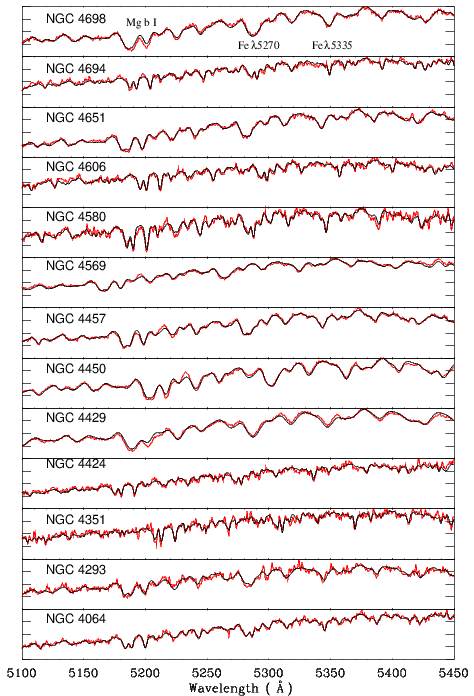}
\end{center}
\caption{ 
Spectra of the center of the sample galaxies and best pPXF model spectra.
Galaxy spectra are
represented in red thick lines,  pPXF model spectra is represented in black solid lines. Clearly are visible Mg b I lines and Fe lines.} 
\label{samplespectra}
\end{figure}

\begin{figure}
\figurenum{4}
\begin{center}
\leavevmode
\epsscale{0.8}
\includegraphics[clip,width=5.5in]{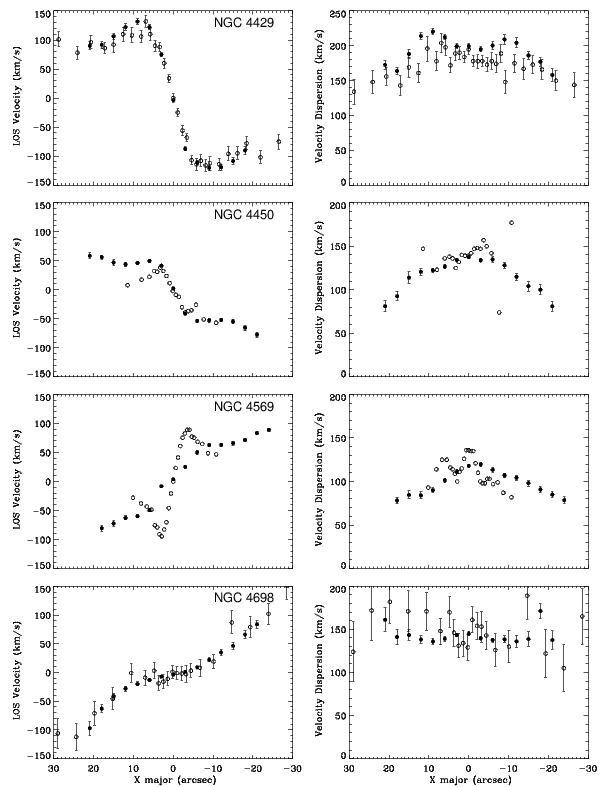}
\caption{ 
Comparison between the stellar kinematics of 5 sample galaxies with the values
taken from the literature.
Solid symbols represent our velocity
determination, whereas open circles represent velocity determination taken from
the literature. From {\em Top} to {\em bottom}: LOS mean velocities and velocity
dispersion for NGC 4429, NGC 4450, NGC 4569 and NGC 4698. NGC 4429 published data
were taken from Simien \& Prugniel (1996), NGC 4450 and NGC 4569 published data
were taken from Fillmore et al. (1986), and NGC 4698 from H\'eraudeau et al. (1999).
}
\end{center}
\label{compplot}
\end{figure}

\begin{figure}
\figurenum{5}
\begin{center}
\leavevmode
\includegraphics[clip,width=5.5in]{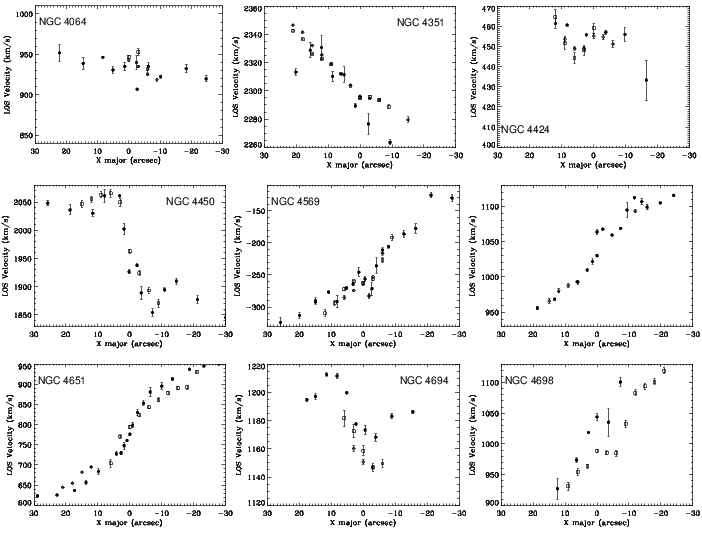}
\caption{
Comparison between the ionized gas kinematics along the major axis for 9 sample galaxies with the values
taken from the literature (Rubin \etal 1999).
Solid symbols represent Rubin's H$\alpha$ velocities,
whereas open diamonds represent our H$\beta$ velocity determination, and open circles our [\ion{O}{3}]
velocity determinations.
}
\end{center}
\label{compvgas}
\end{figure}

The central stellar isovelocity contours indicate rotation plus non-circular motions. 
The S-type twisting in the isovelocity contours  (Fig. $\ref{stellarmapset1}$). 
are consistent with the pattern expected for a strong bar (Fig. $\ref{galaxydspk1}$).
Due to the strong barlike streaming motions, the mean 
stellar kinematic major axis (PA=129$\deg$; PA$\sim$ 100$\deg$ for the central $r=$10") 
is very different from the
photometric major axis of the inner galaxy, which reflects the bar (PA=170$\deg$).
and the photometric major axis of the outer galaxy (PA=150$\deg$), which reflects the disk.

The stellar velocity dispersion field exhibits an off-center peak of 66 $\kms$,
and an asymmetric shape, with the line-widths greater on the NW side than
the SE side. We think this feature is real, although it isn't clear whether it
is a telltale signature of a minor merger, or an uninteresting consequence of
patchy dust extinction.
Overall, the radial velocity dispersion profile is flat, which suggests
some type of gravitational interaction.

The ionized gas emission is confined to the inner 10 arcsec and
forms a bar-like structure (Fig. $\ref{stellarmapset1}$), with
star formation regions (traced by the H$\beta$ emission) along the major
axis of the bar. The ionized gas
(H$\beta$ and [\ion{O}{3}]$\lambda$ 5007 emission), exhibits kinematic
behavior roughly similar to the stars, with barlike streaming motions
(Fig. $\ref{stellarmapset1}$).
Strong CO emission in the central 10$''$ shows even stronger
barlike streaming motions (Cort\'es \etal 2006).\\ 

A more complete analysis can be found in Cort\'es \etal (2006). Its truncated
H$\alpha$ and HI disks (Chung \etal 2009) suggest the action ICM--ISM stripping. Its location
in the outskirts of the cluster ($d_{M87}$ = 2.9 Mpc; Cort\'es \etal 2008), and recent
quenching of the star-formation (just 425 Myr; Crowl \& Kenney 2008), suggest that
this galaxy experienced the combined effect of a gravitational interaction and 
stripping at the outskirts of the cluster.\\

{\bf NGC 4293:} The outer disk of NGC 4293 is tilted with respect to the inner disk, a clear signature of a gravitational interaction.
Optical images shows 2 bright spiral arms which end at $r=$80" in a ring-like feature with PA=75$\deg$.
Beyond this there is a sharp drop in surface brightness, accompanied by a
shift in PA which reaches PA=66 $\deg$ at $r=$ 150".

The central stellar velocity field of NGC 4293 seems largely consistent with rotation. The perturbations in the
isovelocity contours are likely due to the low signal-to-noise ratio of the spectra.
However, the mean stellar kinematic position angle (P.A = 42$\deg$) is very
different from the inner disk optical P.A ($\Psi$ $\sim$ 35 $\deg$), and also different
from the outer optical position angle ($\Psi \sim$ 24$\deg$). (Table $\ref{table4}$).
Analysis of the Spitzer 3.6$\mu$m image shows regular 
isophote shapes with a PA=77$\deg$ in the central $r=$20", bulge-dominated region,
so the inner stellar distribution is aligned with the rest of the galaxy inside $r=$80".
Moreover, there is no clear photometric evidence of a bar, 
although we cannot rule out a bar extended largely along the line-of-sight.
The stellar kinematic major and minor axes are largely perpendicular, so there is also no clear
kinematic evidence of a bar, although a map with higher S/N ratio will provide a better test of this.
We have no other explanation for the large difference in kinematic and photometric PAs.

The stellar velocity dispersion shows a general increase toward the center.
Not all of the structure apparent in the $\sigma$ map is real. Strong dust lanes
near the center may affect the line shapes in some locations.

[\ion{O}{3}] emission (Fig. $\ref{stellarmapset1}$) was only detected
in two regions separated by 10". The scarcity of the ionized gas emission made it
impossible to obtain a reliable extended velocity field.

The misalignment between the kinematical and optical major axes, the disturbed dust lanes and 
tilted outer stellar disk all strongly suggest a gravitational interaction.\\ 
\\

{\bf NGC 4351:} NGC 4351 is lopsided. The continuum light in the central $\sim$~30", 
where there is strong star formation with an irregular pattern,
is significantly offset from the centroid of outer disk isophotes.
Otherwise, there are no obvious disturbances in the light distribution of the outer galaxy,
which lacks star formation. It has been unclear whether the
peculiarities of this galaxy are due to ram pressure or a gravitational interaction.

The surface brightness of the stellar light in the center of NGC 4351
is fainter than most other sample galaxies, so the signal-to-noise ratio 
of the spectra is lower ($\sim$ 20 per pixel in the center).

The stellar velocity field is generally consistent with a rotation pattern
(Fig. $\ref{stellarmapset1}$), except that all the isovelocity contours curve toward the SW.
While not all of the features which deviate from a pure
rotation pattern are real, the unusual curvature  toward the SW may be real.
A similar pattern is observed with high significance in both ionized gas velocity fields (Fig. $\ref{stellarmapset1}$),
as well as the HI velocity field (Chung \etal 2009).
It remains unclear whether this apparently similar disturbance to the stellar and gas velocity fields is due
to a tidal interaction or ram pressure.
While ram pressure doesn't affect stars, much of the stellar light we are measuring in this region
could be from young stars which formed in gas disturbed by ram pressure.
The galaxy's large systemic velocity with respect to the cluster, and
the large projected ICM density at its cluster location
suggest the action of ICM-ISM stripping.

The stellar velocity field displays a mean kinematical P.A of 53$\deg$
$\pm$ 8$\deg$, similar to the optical P.A measured by us at $r$=100"
and very different from the optical P.A of 80$\deg$ given by Koopmann \etal 2001.
Ionized gas velocities (H$\beta$ and [\ion{O}{3}])
exhibit a mean kinematical P.A of about 62$\deg$,
very similar to HI kinematical P.A. given by Chung \etal (2009).

The stellar velocity dispersion over the central field is constant within the errors 
at $\sim$ 25 $\kms$, the lowest value in the sample.
It is difficult to measure dispersion variations in this galaxy, due to both 
the low signal-to-noise ratio, and the fact that 
the measured $\sigma$ ($\sim$ 25 $\kms$) is of the order of the velocity
resolution of the spectra.\\ 

{\bf NGC 4424:} NGC 4424 is a strongly disturbed and highly peculiar galaxy.
It has a heart-shaped stellar light distribution at intermediate radii and shell-like features in outer disk,
clearly indicating a merger or other major gravitational encounter (Kenney \etal 1996; Cort\'es \etal 2006).
Star formation is very strong in the central $\sim$ 1 kpc, but virtually absent at larger radii.
Disturbed dust lanes extend to $\sim$ 3 kpc (Cort\'es \etal 2006).
It has an HI tail extending beyond the main body to the south (Chung \etal 2009),
suggesting either ICM--ISM stripping or a collision with M49.
The precise origin of this combination of peculiar features remains unclear.

The stellar velocity field is largely consistent with a rotation pattern. 
The small velocity range
(400 $\kms$ $\leq V_{los} \leq$ 470 $\kms$) within the inner 25", shows the
remarkably low velocity gradient exhibited by the stellar kinematics,
which was previously observed in the ionized gas by Rubin \etal (1999).

The stellar velocity dispersion field is relatively flat, but shows
three modest peaks slightly offset from the major axis. One is located just north of the nucleus,
and two others are nearly symmetrically located 10" about the nucleus,  roughly at the same location of the
outermost luminous \ion{H}{2} complexes.

Ionized gas intensity line maps show that H$\beta$ and [\ion{O}{3}] emission 
are confined to two star forming regions roughly equidistant from the center ($\sim$ 10"). 
The ionized gas velocity fields are completely different from the stellar velocity field, 
with isovelocity contours roughly perpendicular to the stellar velocity field.
The amount of non-circular motions in the
ionized gas velocity fields made it impossible to obtain reliable gas rotation curves.\\

{\bf NGC 4429:} 
NGC 4429 is an S0 galaxy notable for a very regular dust disk which extends to 10".
It has no other known peculiarity.
Emission lines were not detected.

The stellar velocity field has a well defined spider diagram,
consistent with a rotation pattern (Fig. $\ref{stellarmapset1}$),
except for minor variations along the minor axis to the north.
The isovelocity contours are pinched in the inner 5" due to
a large central velocity gradient.
Two peaks symmetrically located around the
nucleus at 10" suggest a rapidly rotating second stellar component
associated with the dust disk.
Similar features have also
been observed in other early-type disk galaxies such as NGC 4526, NGC 4570 and
NGC 4621 (Emsellem \etal 2004).

The stellar velocity dispersion field exhibits two peaks
offset from the center, and a dip in the central part probably due
by a cold central stellar disk as in
NGC 7332 (Falc\'on-Barroso \etal 2004) and NGC 4526 (Emsellem \etal 2004),
plus the effects of dust extinction on the line profiles.\\

The cold stellar disk embedded with a big bulge could be the result of gas
falling to the center, which can happen due to the presence of gas in a merger (Jesseit \etal 2007).
Since NGC~4429 presently has no gas and no signs of any recent disturbance, we think the merger must have happened a long time ago.\\ 

{\bf NGC 4450:} NGC 4450 is a large spiral with an anemic star-forming disk which appears to be truncated in
the outer galaxy at $r$=60" (Koopmann \etal 2001).
The central region mapped by us includes a bulge and the inner part 
of a large-scale stellar bar with a PA$\sim$10$\deg$  which extends to $r\sim$50".

The stellar velocity field
shows fairly regular kinematics, mostly consistent with a rotation pattern.
A considerable pinching in the
isovelocity contours in the inner 3" indicates a large central velocity gradient.
The two plateaus in the stellar velocity field symmetrically located
at $r\sim$ 5" are similar to the two peaks in NGC 4429, and
suggest the presence of a secondary rapidly rotating component.
The mean central stellar kinematic
position angle differs by -12$\deg$ $\pm$ 5$\deg$ with
respect to the outer optical P.A. 

The stellar velocity dispersion is centrally peaked, and relatively asymmetric. 

H$\beta$ emission is very weak and is confined to only a few fibers within the array.
[\ion{O}{3}]$\lambda$5007 emission is brighter and more extended (Fig. $\ref{stellarmapset1}$).
The [\ion{O}{3}] velocity field seems dominated by rotation, but 
with a position angle offset by 30$\deg$ with respect to the
central stellar velocity field, indicating significant noncircular motions,
or possibly a central gas disk tilted with respect to the stellar disk.
If it is a tilted gas disk, it would probably not survive long due to gravitational torques and gas dissipation, so we think this gas was recently acquired.\\

{\bf NGC 4457:} 
This nearly face--on galaxy has significant HI and star formation in the central 30", and virtually nothing beyond.
The inner ISM-rich part of the galaxy is dominated by one peculiar strong spiral arm.

The stellar velocity field displays a pattern consistent
with rotation. The stellar kinematic P.A. is 70$\deg$ and
probably corresponds to the line-of-nodes of the galaxy. 

As in NGC 4429 and NGC 4450, there are
pinched isovelocity contours within the inner 5", and
two peaks in the stellar velocity located at around 5".

The stellar velocity dispersion is relatively constant over the 
central field, although has a modest peak in the very center.
The azimuthally averaged sigma profile shows 
a possible secondary peak at r=10", although the irregular pattern
in the sigma map, and the low signal-to-noise ratio make this uncertain.

The ionized gas intensity maps (Fig. $\ref{stellarmapset1}$) display emission
near the nucleus and on the spiral arms. The ionized gas velocity fields
show a dominant rotational component, but also large non-circular motions
indicating some type of disturbance.
The isovelocity contours appears curved systematically toward the West, although our map
is sparse. However, this curvature
is also present in the more extended and detailed H$\alpha$ velocity field presented in Chemin \etal (2006).
Both maps show a strongly blueshifted region close to the core of the galaxy.
The kinematic PA of the ionized gas is close to that of the stars, 
but the H$\beta$ and [\ion{O}{3}] velocities are systematically lower than the stellar
velocities by 30--40 $\kms$. Along the stellar kinematic minor axis,
ionized gas velocities show big differences with respect to the stars,
becoming systematically redshifted on both sides of the nucleus,
indicating non-circular motions in the gas.
This gas isovelocity pattern resembles that observed in NGC 4351.
We were unable to obtain reliable rotation curves
for the gas due to the amount of non-circular motions in the velocity fields.

The gas kinematics seem consistent with ongoing ICM-ISM stripping.
In fact the presence of a peculiar H$\alpha$ arm could
also be explained as the result ICM wind (Schulz \& Struck 2001).\\

{\bf NGC 4569:} 
This large spiral has an HI and star-forming disk truncated at r$\sim$90", and an
unusual extraplanar arm of gas \& star formation due to ram pressure stripping.
Its outer stellar disk ($r \sim$170"--270"; PA$\sim$25 $\deg$) is slightly
twisted with respect to inner disk ($r \sim$ 70--140"; PA$\sim$20$\deg$).
It has a bright nucleus, is a known LINER (e.g. Keel 1996), and has a nuclear outflow.

The central stellar velocity field is largely consistent with a rotation pattern, although
velocities on NW side of the minor axis are systematically
blueshifted by a few $\kms$ with respect to pure rotation, perhaps due to
the effects of the prominent dust lanes present in this galaxy
(Fig. $\ref{galaxydspk1}$), since they are strong on the NW side.
In the central 10", there are other apparent deviations from pure rotation 
without any clear pattern, and these might also be caused by the effects of dust.
This galaxy exhibits a small difference between the mean stellar kinematic PA (32$\deg$)
and the optical photometric PA (23$\deg$)($\Delta P.A \sim 9\deg$), which could be
the result of a tidal interaction.

The stellar velocity dispersion has a strong central peak, and drops sharply with radius.
The drop off is steeper along the minor axis than the major axis, 
a reflection of the highly inclined viewing angle.

The ionized gas intensity maps (Fig. $\ref{stellarmapset1}$) show that both H$\beta$
and [\ion{O}{3}] emission are associated with the central starburst.
H$\beta$ but not [\ion{O}{3}] is also located to the NW where the star formation ring is located, 
and [\ion{O}{3}] but not H$\beta$ is detected south of the nucleus.

The gas velocity fields are very disturbed, indicating significant non-circular motions.
Along the minor axis the gas velocities reach a maximum value of -290 $\kms$, 
with a velocity amplitude of about 70 $\kms$. These are likely due to the nuclear outflow known to exist from the extended radio and H$\alpha$ features along the minor axis (Chy\.{z}y \etal 2006).

We couldn't obtain ionized gas rotation curves due to the amount of
large non-circular motions in the gas.\\

{\bf NGC 4580:}
NGC 4580 has HI and H$\alpha$ emission only out to a radius of 26",
where there is a prominent star formation ring. The outer disk has stellar spiral arms.

The stellar velocity field displays a nice spider diagram consistent with a 
rotation pattern (Fig. $\ref{stellarmapset2}$), and shows no significant
non-circular motions.
The central stellar kinematic major axis is very close to the photometric major axis. 

The stellar velocity dispersion is relatively constant over the central field, and
the irregular features in the dispersion map are probably not real.

The ionized gas intensity maps exhibit H$\beta$ emission over much of the array
(Fig. $\ref{stellarmapset2}$), with the strongest emission associated with
the star formation ring. The [\ion{O}{3}] emission is poorly correlated with
the H$\beta$ emission, 
with peaks in regions where H$\beta$ is weak.

The H$\beta$ velocity field shows an overall pattern of rotation with 
the same kinematic position angle as the stars.  The gas isovelocity 
contours show wiggles indicating localized non-circular motions
which may be associated with the star forming ring.

The undisturbed, stellar and ionized gas kinematics 
indicates that there are no signatures
of a gravitational interaction or ongoing strong ICM-ISM stripping.
The presence of a truncated
H$\alpha$ disk  with stellar spiral arms in a gas poor outer disk
indicates that ICM-ISM stripping probably acted within the last Gyr (475 Myrs ago;
Crowl \& Kenney 2008).\\

{\bf NGC 4606:} 
NGC 4606 has a truncated/compact H$\alpha$ morphology, 
meaning there is strong star formation in the central ~1 kpc, but none beyond.
It has a disturbed stellar body with non-elliptical isophotes, indicating
a strong gravitational encounter. It is an apparent pair with the spiral NGC 4607,
separated by 17 kpc and 600 $\kms$.

The stellar velocity field is dominated by a rotation pattern (Fig. $\ref{stellarmapset2}$),
but some minor distortions are present. 
The stellar kinematical major axis P.A. = 39$\deg$ in the inner 20" agrees well with 
the PA=44$\deg$ measured from optical photometry of the outer galaxy.

The stellar velocity dispersion is essentially constant, with
a value of 49 $\pm$ 3 $\kms$. Two apparent peaks in $\sigma$
are located about 12" N and S from the center, although the
significance of these features is low.
Simulations show that flat velocity dispersion profiles (Bendo \& Barnes 2000)
and off-axis $\sigma$ peaks (Jesseit \etal 2007)
can be produced in mergers
with mass ratios $\sim$3:1.

The ionized gas emission
is confined to the inner 10" (Fig. $\ref{stellarmapset2}$). The strongest
emission arises from an elongated bar-like string of luminous \ion{H}{2} complexes,
similar to those in the other truncated/compact galaxies,
NGC 4064 and NGC 4424. There is a westward extension in [\ion{O}{3}] emission
with no H$\beta$ counterpart, along the minor axis.
The H$\beta$ velocity field exhibits a gradient along the major axis which could be
consistent with rotation. Ionized gas emission is too scarce to derive further conclusions.\\

{\bf NGC 4651:} 
Over most of the optical disk NGC 4651 looks like a relatively normal spiral galaxy, but
the outer parts are strongly disturbed, with shell-like features, and a peculiar straight tail (Chung \etal 2009), probably due to a minor merger.

The central stellar velocity field is well-order with a pattern strongly dominated by rotation.
The stellar kinematical major axis P.A. =82$\deg$ in the inner 20" agrees well with 
the PA=80$\deg$ measured from both the optical photometry and the 
HI velocity field of Chung \etal (2009) at intermediate radii ($r \sim$ 30--90").
While to align the DensePak array we used PA=71$\deg$,  from the optical 
photometric major axis measured at $R_{\rm 25}$ (Koopmann \etal 2001), this PA is characteristic
of the outer galaxy ($r>$100"), where the galaxy appears disturbed.

The stellar velocity dispersion is centrally peaked,
and decreases strongly with radius.  A clear elongation is detected with high significance
along the major axis of the galaxy. This feature is likely
associated with a deviation from the gaussian distribution in the LOSVD, which contains
a strong $h_{3}$ component (see \S 5.3). 

This is just inside the radius of the unusually bright, tightly wound spiral arms at $r \sim$20$''$,
suggesting that both might be related to the minor merger.

The H$\beta$ intensity map exhibits emission from star formation in the tightly wound spiral arms at $r \sim $20$''$
(Fig. $\ref{stellarmapset2}$),
but very weak emission from the center. The [\ion{O}{3}] intensity map shows
strong emission over the nucleus and the spiral arms. 
Both ionized gas velocity fields appear largely similar to the stellar
velocity field beyond the central 10",  although
both H$\beta$ and [\ion{O}{3}] maps show wiggles in the isovelocity contours 
which could be associated with spiral arms.
Within the inner 10",  the [\ion{O}{3}] map exhibits a twisting in the
isovelocity contours, indicating non-circular motions in this gas.\\

{\bf NGC 4694:} 
NGC 4694 has strong emission from the nuclear region, due to star formation
and/or an AGN, and otherwise it has weak star formation in the central kpc, and none beyond.
A disturbed complex of HI extends across the minor axis to the nearby faint
galaxy VCC 2062, indicating some kind of gravitational interaction, perhaps
a minor merger or gas accretion event (Chung \etal 2009).

The stellar velocity
field is largely consistent with a rotation pattern, although shows some
modest apparent deviations from circular motions which have low significance.
Along the minor axis (P.A=56$\deg$), the stellar velocities do not display any gradient.
The P.A. of the stellar kinematical major axis in the center agrees well with 
the P.A. measured from optical photometry in the outer disk.

The stellar velocity dispersion is practically flat with a small dip in the center. Some of
the apparent variations exhibited in the outer parts are likely
artifacts due to a low signal-to-noise ratio.

The ionized gas intensity maps show emission within the central 10".
The nucleus is relatively strong, and there are fainter features
extending to the NW and SW of the nucleus (Fig. $\ref{stellarmapset2}$).
The ionized gas velocity fields appear disturbed.
 Along the major axis, ionized gas velocities are
systematically lower by 25 $\kms$ than the stars, and in places show large
velocity gradients.\\

{\bf NGC 4698:} 
This remarkable galaxy NGC 4698 has a bulge that rotates orthogonally with respect to the disk
(Bertola \etal 1999). The disk appears undisturbed, suggesting that this galaxy is the product of
an ancient merger.

The stellar velocity field shows at least two
distinct kinematical components (Fig. $\ref{stellarmapset2}$). The inner
10" exhibits S-shaped isovelocity contours near the minor axis, due to the
orthogonally rotating core. 
The effect is less evident that in previously published data (Bertola \etal 1999), due to the
smoothing effect introduced by the width of the fibers.
In the outer 10", the velocity field is
consistent with a rotation pattern in most respects, 
except that the isovelocity contours are concave.
These concave contours mean that the line-of-sight velocity increases
as we going outwards from the major axis, which is probably 
due to the effect of the orthogonally rotating bulge.
This galaxy exhibits almost no stellar rotation within the inner 10".
Beyond the inner 10", the LOS stellar velocities rise dramatically,
due to the increasing influence of the stellar disk.

The stellar velocity dispersion has a modest peak at the center, but
otherwise remains flat along the major and minor axes.

H$\beta$ emission was not detected. On the contrary, [\ion{O}{3}] emission is very strong in the center
and detected over the entire array (Fig. $\ref{stellarmapset2}$). The [\ion{O}{3}]
velocity field is well-ordered, but very different from the stellar velocity field.

The pattern seems largely consistent with rotation, although with a PA which changes with increasing radius,
from 120$\deg$ at the center to 150$\deg$ at 25". 
(In order to produce the  [\ion{O}{3}] rotation curve in figure $\ref{rotcurvset1}$, we used rings with different P.As.)
In the inner $r \sim$ 5", the [\ion{O}{3}] kinematic PA is roughly perpendicular to the local stellar kinematic PA.
At $r \sim$25", the PAs are offset by $\sim$20 $\deg$.
We suspect that the gas distribution is non-planar, as the
equilibrium plane of the gas is expected to gradually change from larger radii, 
where the outer disk dominates the gravitational potential, to smaller radii,
where the orthogonally rotating bulge dominates the potential.

\newpage


\clearpage

\begin{figure}
\figurenum{6}
\begin{center}
\leavevmode
\includegraphics[width=6.65in,clip]{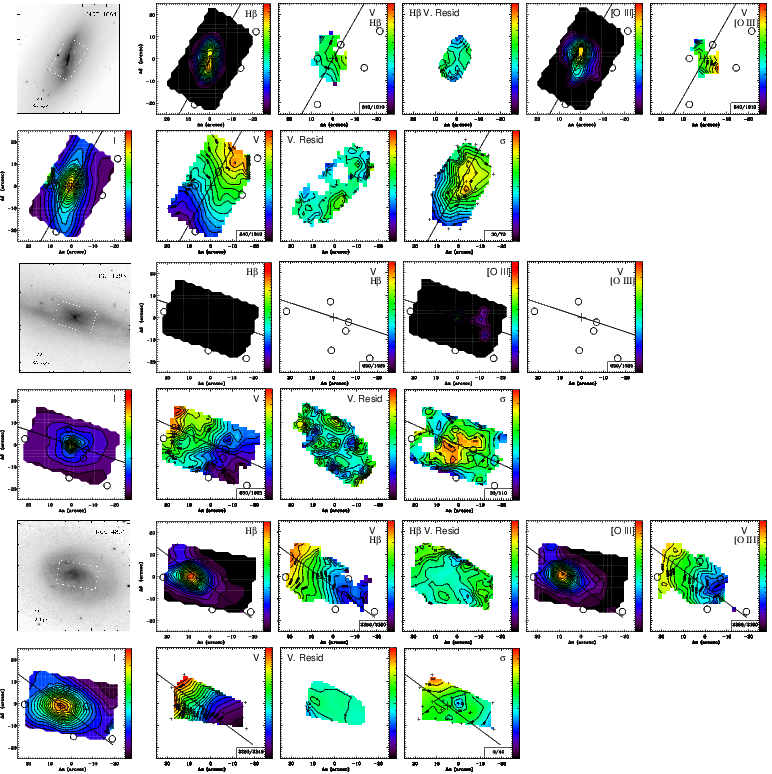}
\end{center}
\caption{
Maps of the stellar and ionized gas kinematics of the sample galaxies.
Solid lines
represent the optical P.A of the galaxies. Crosses represent the
position of the peak in the continuum map,
open circles the dead
fibers. The white rectangular box represents the area covered by DensePak.
In the maps, North is up and East left. For each galaxy we display (first row), R-band
image and where available H$\beta$ intensity map, H$\beta$ velocity field, H$\beta$ velocity residuals
, [\ion{O}{3}]$\lambda\lambda$
5007 {\AA} intensity map, [\ion{O}{3}]$\lambda\lambda$ 5007 {\AA} velocity field,
[\ion{O}{3}]$\lambda\lambda$ 5007 {\AA} velocity residuals. Second row; stellar reconstructed continuum map,
stellar velocity field, stellar velocity residuals
stellar dispersion field, and where available $h_{\rm 3}$ and $h_{4}$ moment maps.
}
\label{stellarmapset1}
\end{figure}

\clearpage

\begin{figure}
\figurenum{6}
\begin{center}
\leavevmode
\vbox{
\includegraphics[width=6.65in,clip]{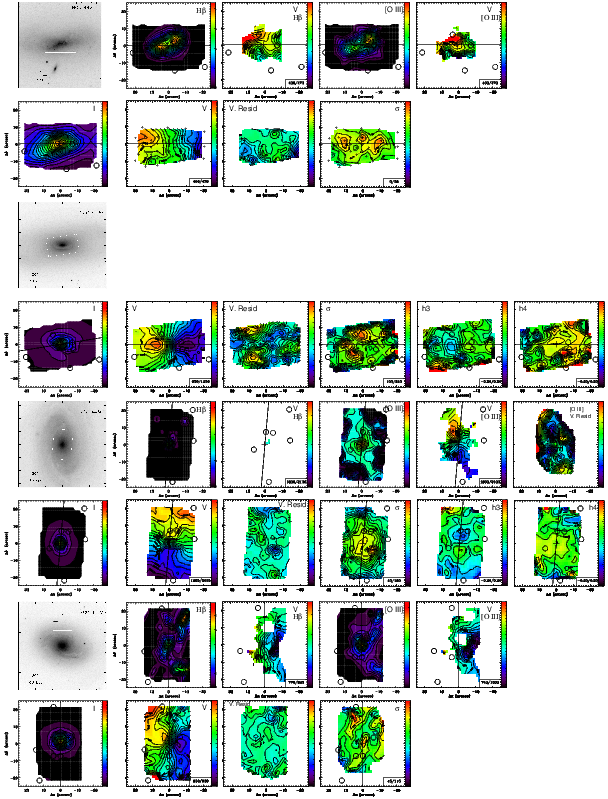}
}
\end{center}
 \caption{{\em Continued}
}
\label{stellarmapset2}
\end{figure}

\begin{figure}
\figurenum{6}
\begin{center}
\leavevmode
\vbox{
\includegraphics[width=6.65in,clip]{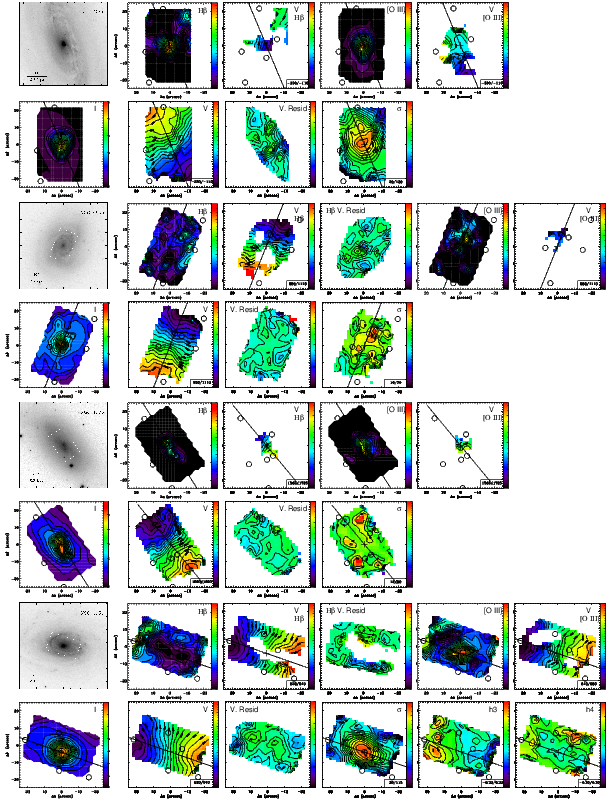}
}
\end{center}
 \caption{{\em Continued}
}
\end{figure}

\begin{figure}
\figurenum{6}
\begin{center}
\leavevmode
\vbox{
\includegraphics[height=4.3in,clip]{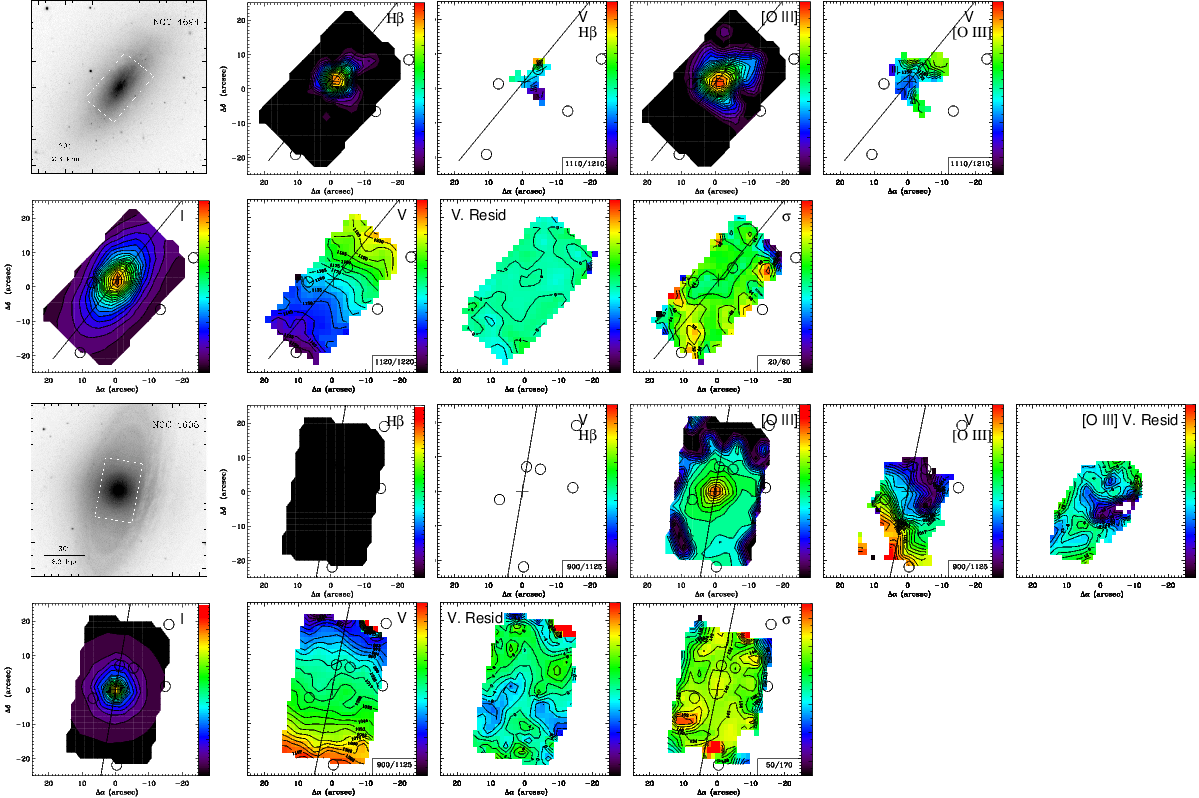}
}
\end{center}
 \caption{{\em Continued}
}
\end{figure}

\begin{figure}
\figurenum{7}
\begin{center}
\leavevmode
\vspace{0.5cm}
\vbox{
\hbox{
\includegraphics[width=2.85in,clip]{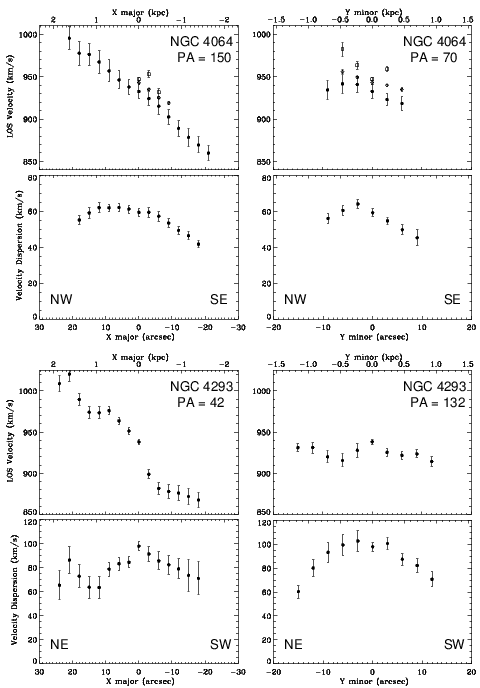}
\includegraphics[width=2.85in,clip]{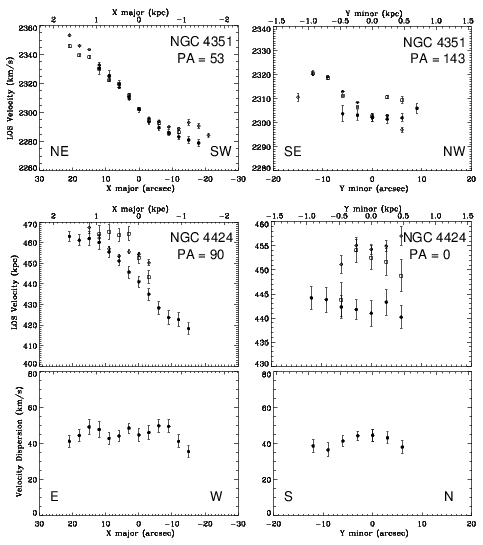}
}
\hbox{
\includegraphics[width=2.85in,clip]{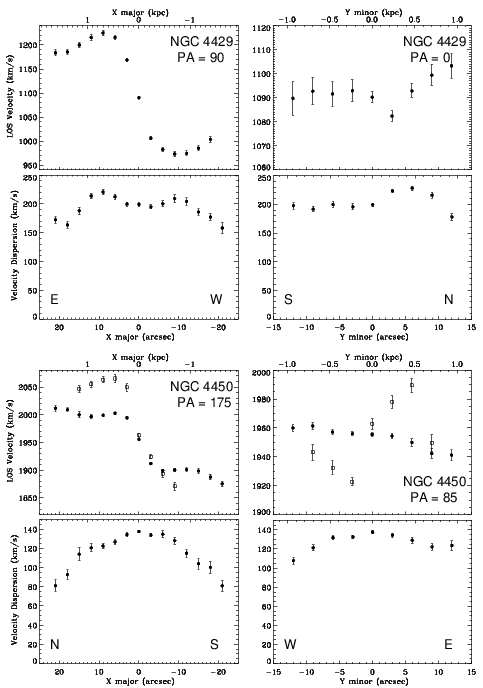}
\includegraphics[width=2.85in,clip]{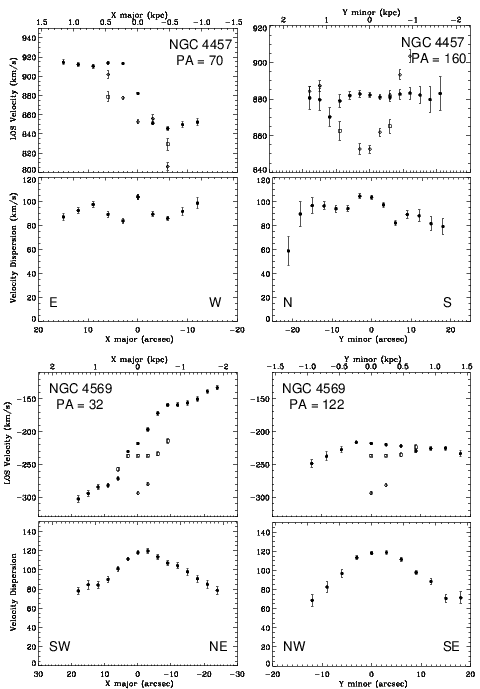}
}
}
\end{center}

\vspace{-1.0cm}
\caption{
Stellar and ionized gas kinematics along kinematical major and minor axes.
Solid symbols represents
stellar kinematics, H$\beta$ gas kinematics is represented by diamonds, and [\ion{O}{3}] gas kinematics is represented by
squares.
}
\label{slicesvel1}
\end{figure}

\begin{figure}[ht]
\figurenum{7}
\begin{center}
\leavevmode
\vbox{
\hbox{
\includegraphics[width=3.5in,clip]{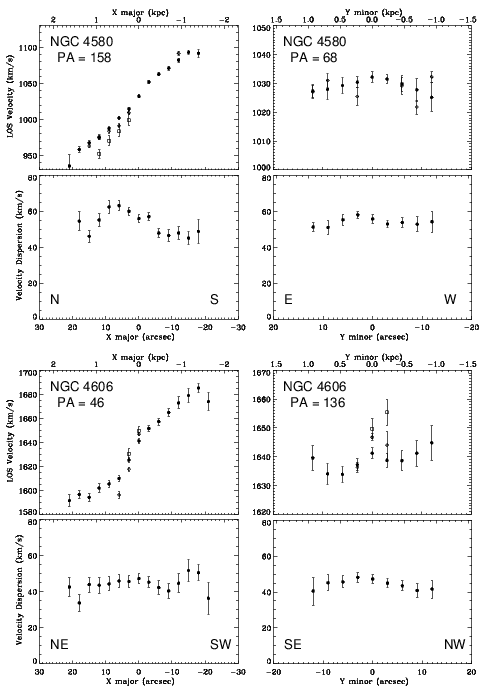}
\includegraphics[width=3.5in,clip]{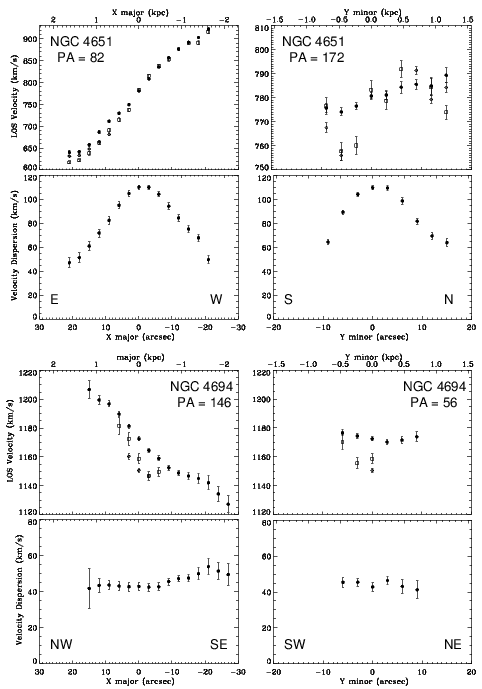}
}
\includegraphics[width=3.5in,clip]{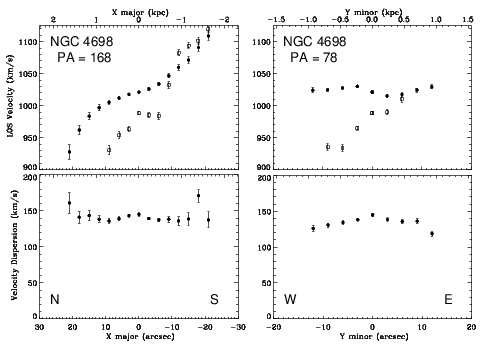}
}

\end{center}

\caption{
{\em Continued}}
\label{slicesvel2}
\end{figure}


















\begin{figure}[ht]
\figurenum{8}
\begin{center}
\leavevmode
\includegraphics[width=6.0in,clip]{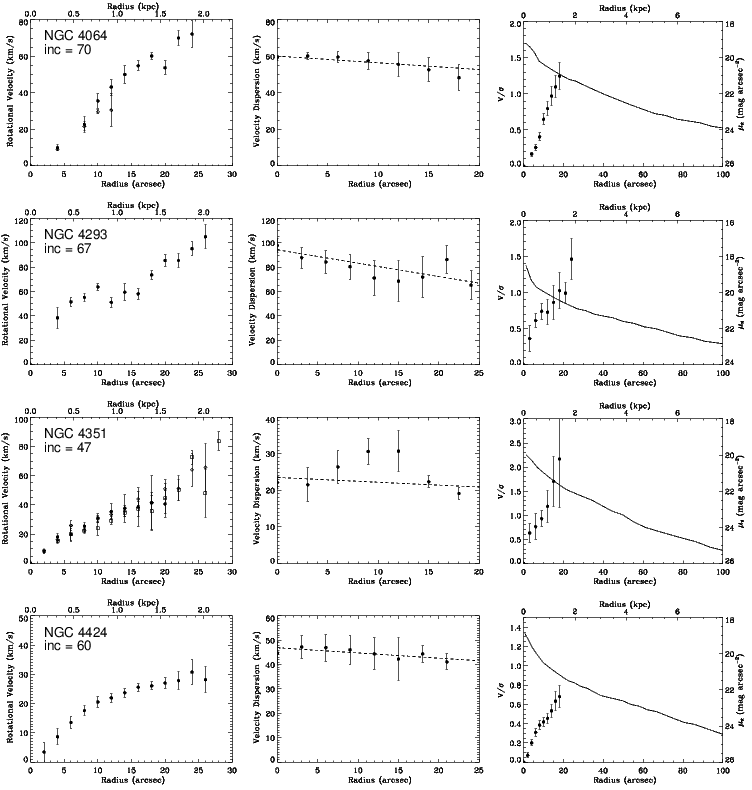}
\end{center}

\caption{
Stellar and ionized gas rotation curves, velocity dispersion profiles, and $V/\sigma$ ratios.
{\em Left panels:} Stellar and ionized gas rotation curves. Solid symbols represent stellar velocities,
diamonds represent H$\beta$ velocities, and squares represent [\ion{O}{3}] velocities. {\em Middle panels:}
Mean stellar velocity dispersion profiles. Solid symbols represent stellar velocity dispersion, and
dashed lines represent the best linear profile fitted to stellar velocity dispersions. {\em Right panels:} Ratio between the stellar rotation curve and the stellar velocity dispersion
profile. Open and solid symbols represent different sides of the galaxy. Solid line represents the R-band radial light profile from Koopmann \etal 2001.}
\label{rotcurvset1}

\end{figure}

\begin{figure}[ht]
\figurenum{8}
\begin{center}
\leavevmode
\includegraphics[width=6.0in,clip]{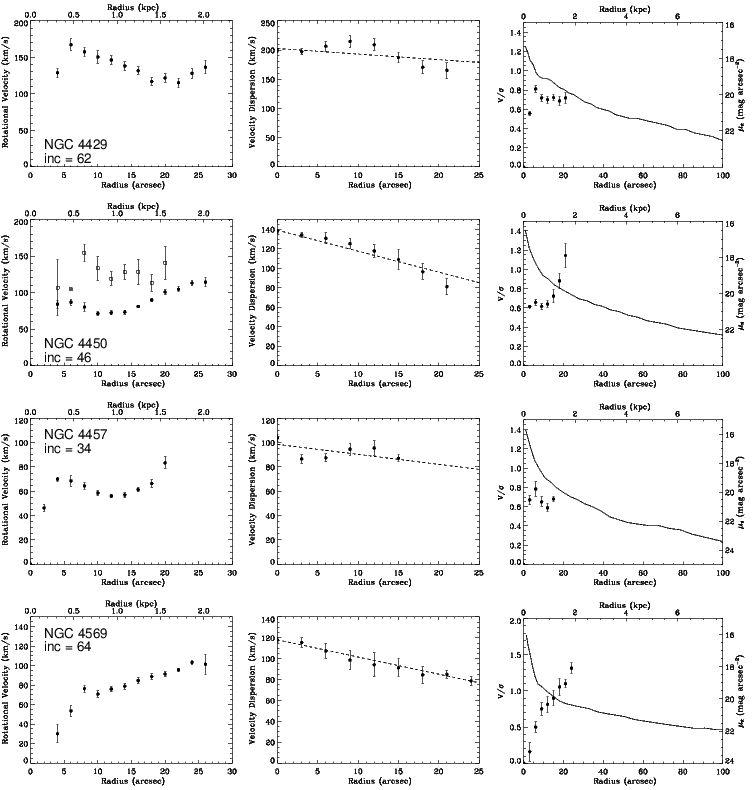}
\end{center}
\caption{
{\em Continued}
}
\label{rotcurvset2}
\end{figure}

\begin{figure}[ht]
\figurenum{8}
\begin{center}
\leavevmode
\includegraphics[width=6.0in,clip]{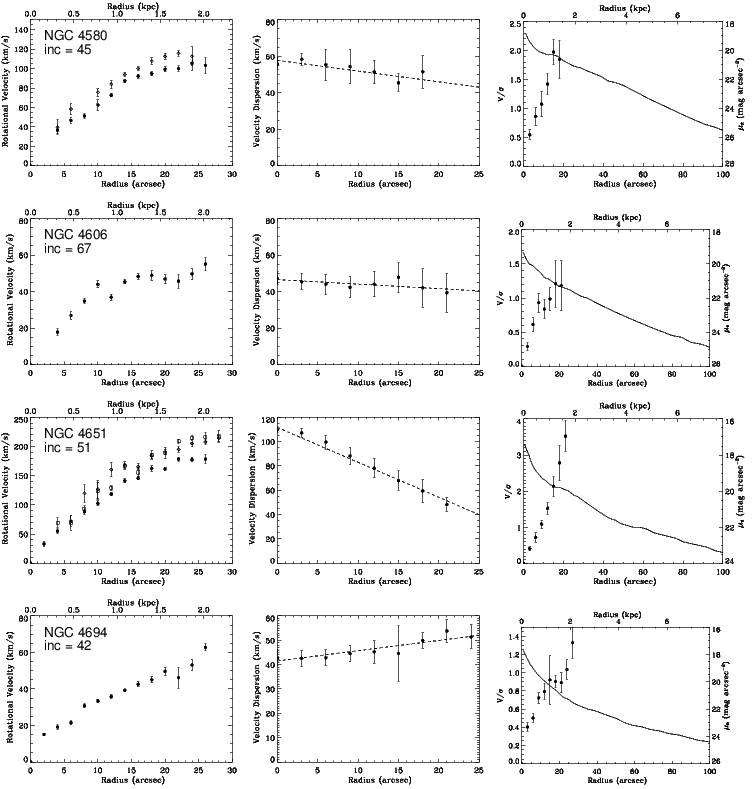}
\includegraphics[width=6.0in,clip]{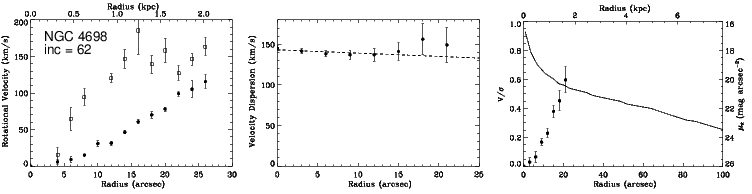}
\end{center}
\caption{
{\em Continued}
}
\label{rotcurvset3}
\end{figure}

\clearpage



\begin{figure}[ht]
\figurenum{9}
\begin{center}
\vbox{
\includegraphics[height=2.45in,clip]{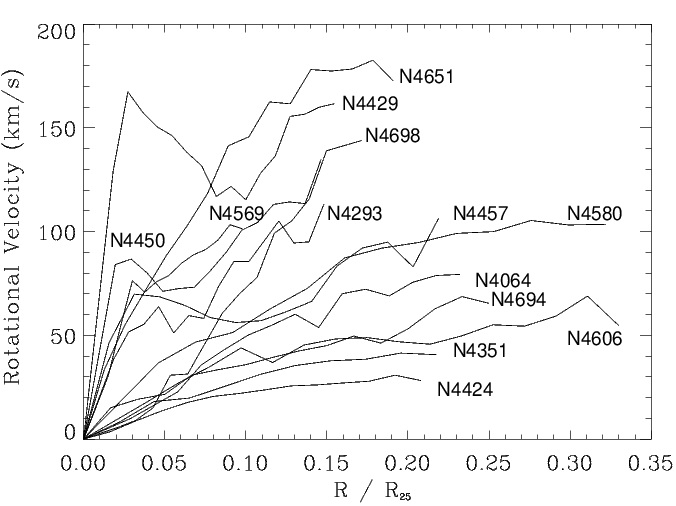}
\includegraphics[height=2.45in,clip]{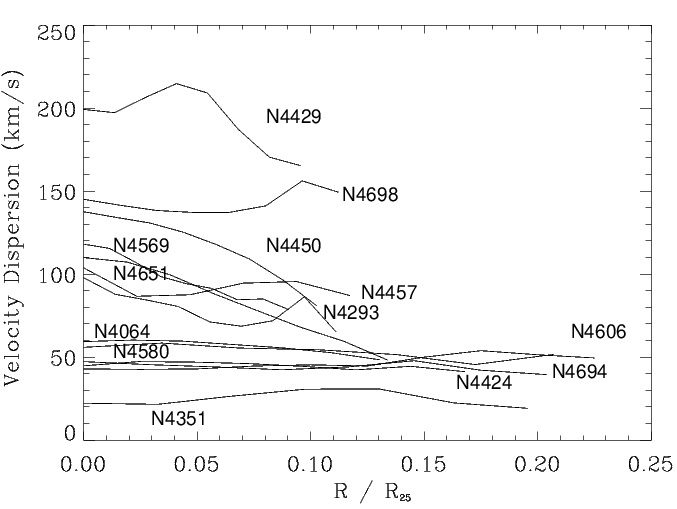}
\includegraphics[height=2.45in,clip]{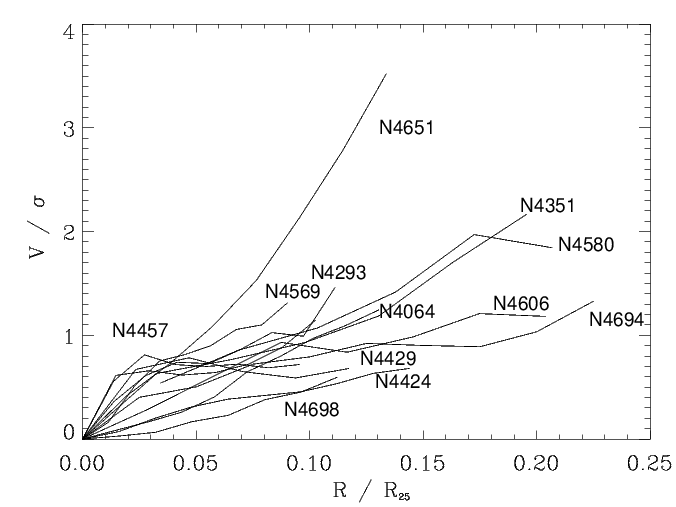}
}
\end{center}
\caption{
Stellar mean rotation curves, mean velocity dispersion profiles, and mean $V/\sigma$ ratios.
All sample galaxies are represented in these plots. Galactocentric radii for sample galaxies were normalized
by $R_{25}$ (Koopmann et al. 2001) defined as the radius were the surface brightness of the galaxy is
25 mag arcsec$^{-1}$ in R-band.}
\label{samplekin}
\end{figure}

\begin{figure}[ht]
\figurenum{10}
\begin{center}
\leavevmode
\includegraphics[width=5.0in,clip]{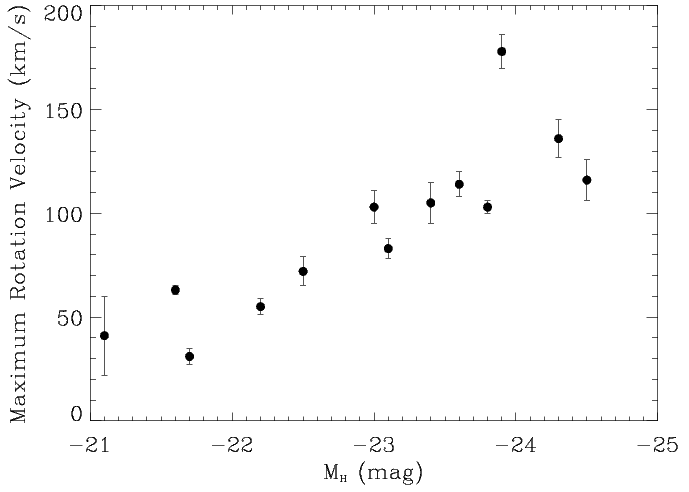}
\includegraphics[width=5.0in,clip]{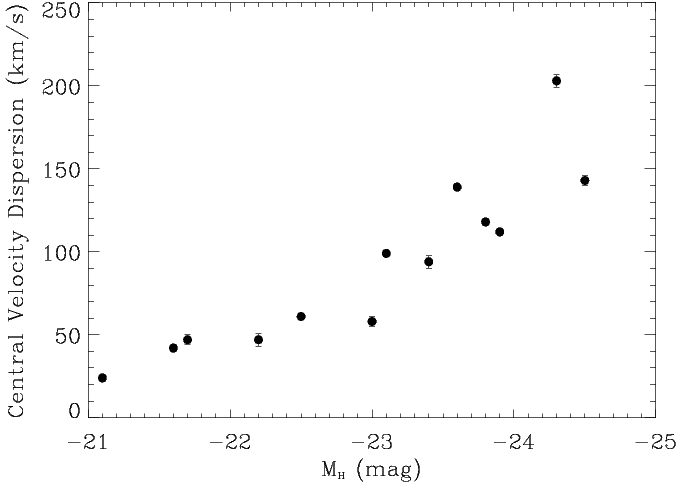}
\end{center}
\caption{
Maximum stellar velocity amplitudes and galaxy absolute magnitudes.
The maximum stellar velocities are correlated with the brightness of the
galaxies. {\em Top:} Maximum stellar rotation velocities and $M_{H}$ magnitudes
{\em Bottom:} Central stellar velocity dispersion and $M_{H}$ magnitudes.
Velocities appear clearly correlated with $M_{H}$, indicating that the
amplitude differences are due to differences in mass.}
\label{correl}
\end{figure}

\begin{figure}[ht]
\figurenum{11}
\begin{center}
\leavevmode
\includegraphics[width=5.5in,clip]{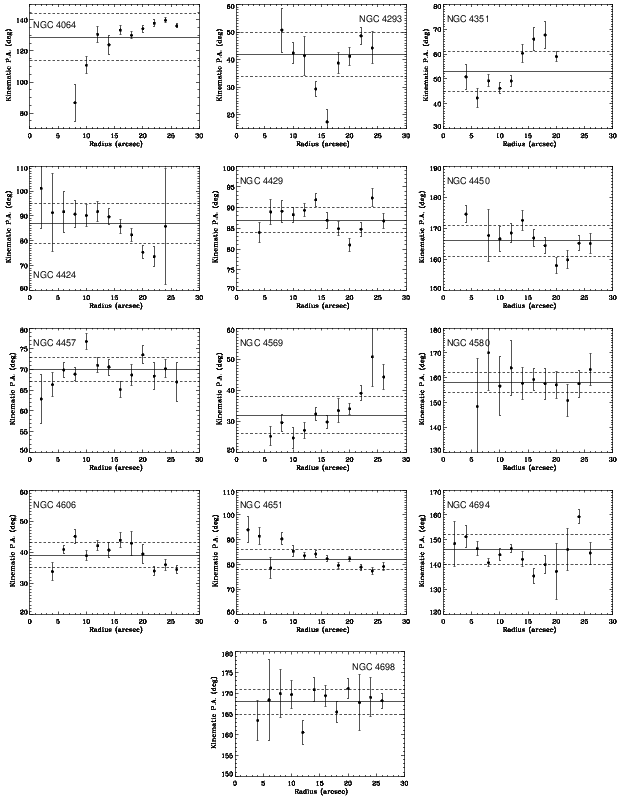}
\end{center}
\caption{
Stellar kinematic PA as function of radius for sample galaxies. Solid symbols represent the stellar kinematic position angles derived from the tilted ring models (section 3.3). Solid lines represent the mean kinematic P.A. The standard deviations of the mean are represented by dashed lines.}
\label{paset}
\end{figure} 


\newpage

\begin{figure}[ht]
\figurenum{12}
\begin{center}
{
\vbox{
\includegraphics[height=2.40in,clip]{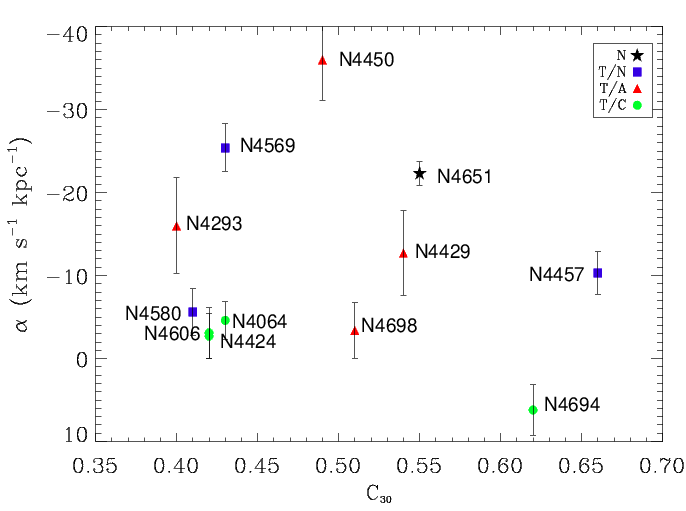}
\includegraphics[height=2.40in,clip]{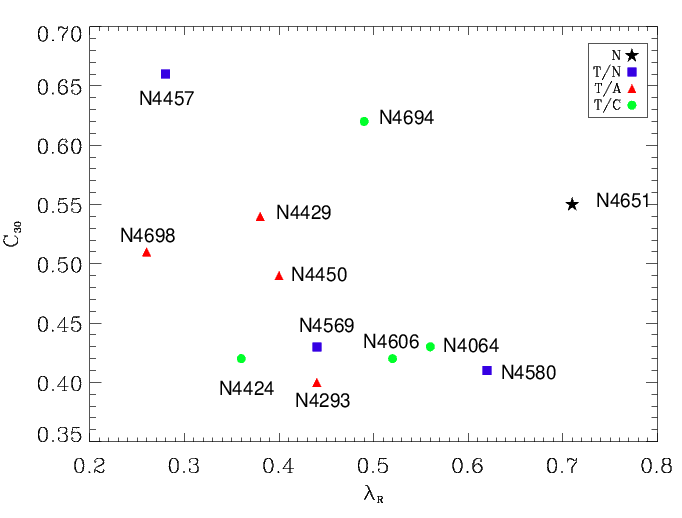}
\includegraphics[height=2.40in,clip]{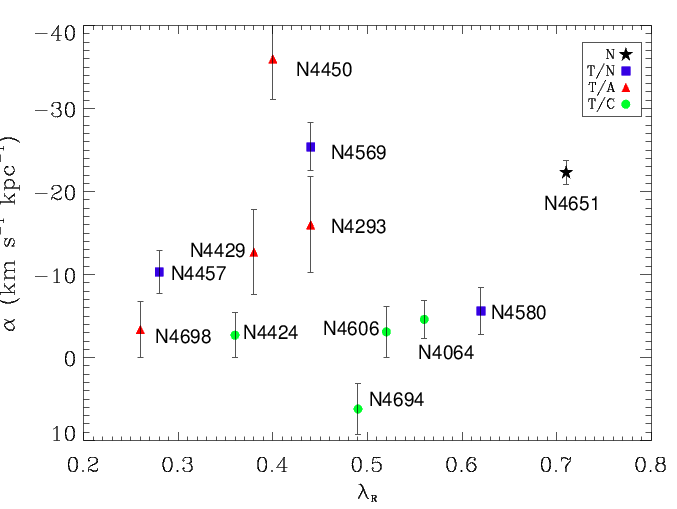}
}
}
\end{center}
\caption{
$\lambda_{\rm R}$ parameter, concentration ($C_{30}$) and velocity
dispersion slope $\alpha$.
{\em Top panel}: $C_{30}$ and $\alpha$. {{\em Middle panel}}:  $C_{30}$ and $\lambda_{\rm R}$. {{\em Bottom panel:}}
$\lambda_{\rm R}$ and $\alpha$. 
Symbols represent different star formation classifications: 
black stars = Normal (N), 
red triangles = Truncated/Anemic (T/A),
blue squares = Truncated/Normal (T/N), and 
green circles =  Truncated/Compact (T/C).
Here we see that
galaxies with star formation classes T/A and T/N span a wide range of $\alpha$ values,
but T/A tends to have low $\lambda_{R}$ ($<$ 0.45). T/C galaxies
have a nearly flat velocity dispersion profiles.}
\label{dyncorr}
\end{figure}

\newpage

\begin{figure}[ht]
\figurenum{13}
\begin{center}
\leavevmode
{
\includegraphics[width=6.0in,clip]{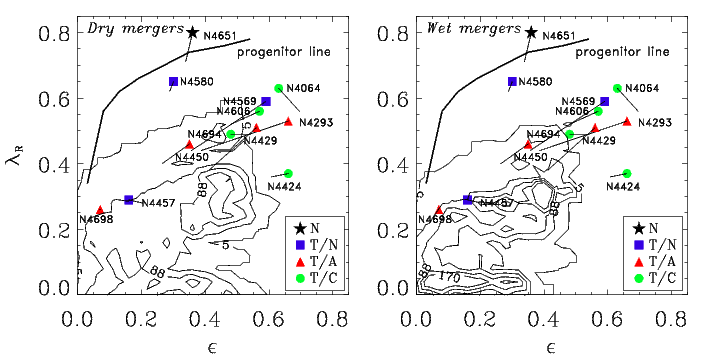}
}
\end{center}
\caption{
Distribution of merger remnants with mass ratios of 3:1 to 1:1 in the $\lambda_{R}$--$\epsilon$
plane. 
Symbols represent different star formation classifications: 
black stars = Normal (N), 
red triangles = Truncated/Anemic (T/A),
blue squares = Truncated/Normal (T/N), and 
green circles =  Truncated/Compact (T/C).
Solid lines connect the uncorrected (observed) and corrected (to R$_{\rm 1/2}$)
values of $\epsilon$ and $\lambda$$_{\rm R}$; the symbols are plotted at the corrected values. Contours represent the distribution of 1:1 and 3:1 merger remnants from Jesseit et al. (2009)
simulations. The progenitor galaxy location is indicated as a
thick line. 
The left panel displays collisionless mergers (dry), and the right panel displays mergers remnants
which formed with gas (wet). Only two galaxies are close to the progenitor line (NGC 4580 and NGC
4651), so have not experienced any major gravitational disturbance. The remaining galaxies are dynamically warm or lukewarm,
thus have experienced significant gravitational disturbances.
}
\label{mergers}
\end{figure}

\newpage

\begin{figure}[ht]
\figurenum{14}
\begin{center}
\leavevmode
{
\hbox{
\includegraphics[width=3.0in,clip]{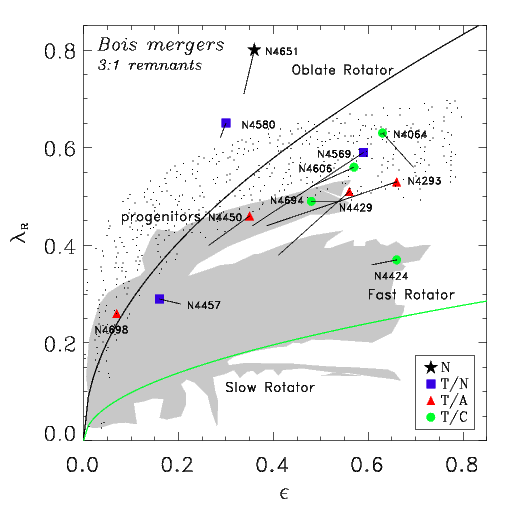}
\includegraphics[width=3.0in,clip]{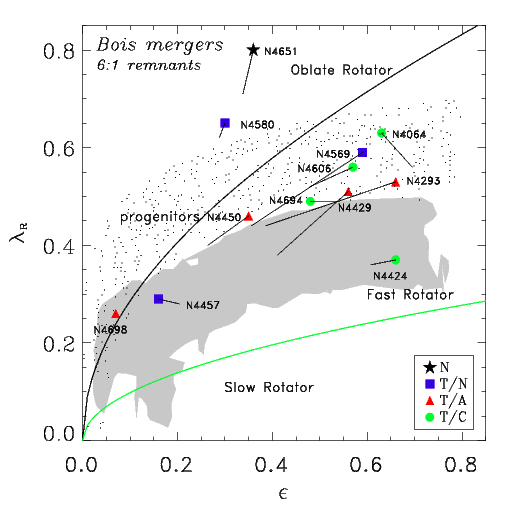}
}
}
\end{center}
\caption{
Sample galaxies in the $\lambda_{R}$--$\epsilon$ plane. 
Symbols represent different star formation classifications: 
black stars = Normal (N), 
red triangles = Truncated/Anemic (T/A),
blue squares = Truncated/Normal (T/N), and 
green circles =  Truncated/Compact (T/C).
Solid lines connect the uncorrected (observed) and corrected (to R$_{\rm 1/2}$) values of $\epsilon$ and $\lambda$$_{\rm R}$; the symbols are plotted at the corrected values. Black
solid line represents the location of oblate rotators, green line represents the boundary between slow and
fast rotators. Black dots represent spiral-like progenitor galaxies from the Bois et al. (2011) simulations. 
The grey shading covers the region of the Bois et al. (2011) merger remnants. The
left panel displays the merger remnant locations with a mass ratio of 3:1. Right panel displays the merger
remnant locations with a mass ratio of 6:1. 
}
\label{mergersbois}
\end{figure}

\newpage

\clearpage
\begin{turnpage}
\begin{deluxetable}{ccccccccccccccc}
\tabletypesize{\scriptsize}
\setlength{\tabcolsep}{0.05in} 
\tablecolumns{12}
\tablewidth{0pc}
\tablecaption{Galaxy Sample Properties\label{table1}}
\tablehead{
\colhead{}  &  \colhead{}  & \colhead{} & \colhead{}  & \colhead{}  & \colhead{}  & \colhead{}
 & \colhead{ } & \colhead{} & \colhead{R$_{25}$} & \colhead{Inc.} & \colhead{P.A} & \colhead{V$_{helio}$} & \colhead{D$_{M87}$} & \colhead{D} \\
\colhead{Name} & \colhead{R.A. (J2000)} & \colhead{Decl. (J200)} & \colhead{RSA/BST}
& \colhead{RC3} & \colhead{C$_{30}$} & \colhead{SFC} & \colhead{H$_{c}$} & \colhead{M$_{H}$} & \colhead{(arcsec)} & \colhead{(deg)}
& \colhead{(deg)} &
\colhead{($\kms$)} & \colhead{(deg)} & \colhead{(Mpc)} \\
\colhead{(1)} & \colhead {(2)} & \colhead{(3)} & \colhead{(4)} & \colhead{(5)} &
\colhead{(6)} & \colhead{(7)} & \colhead{(8)} & \colhead{(9)} & \colhead{(10)} &
\colhead{11} & \colhead{12} & \colhead{13} & \colhead{14} & \colhead{15} }
\startdata
NGC 4064 & 12 04 11.2 &  18 26 36 & SBc(s): &    SB(s)a:pec  & 0.43 & T/C  &  8.78\tablenotemark{a} & -22.5 $\pm^{0.1}_{0.2}$ & 138 & 70
 & 148 & 931 $\pm$  12 & 8.8 & 18.0 $\pm^{1.3}_{0.8}$ \\
NGC 4293 & 12 21 12.8 &  18 22 57  &   Sa pec  &   (R)SB(s)0/a & 0.40 & T/A  &  7.35 & -23.4 $\pm^{0.3}_{0.2}$ & 216 &
67 & 62 & 930 $\pm$ 14 & 6.4 & 14.1 $\pm^{1.0}_{1.5}$\\
NGC 4351 & 12 24 01.6 &  12 12 18  &   Sc(s) II.3 &SB(rs)ab: pec & 0.37 & T/N[s] & 10.44 & -21.1 $\pm$ 0.1 & 92 &
 47 & 65 & 2317 $\pm$ 16 & 1.7 & 20.3 $\pm^{0.8}_{1.2}$ \\
NGC 4424 & 12 27 11.5 &  09 25 15  &   Sa pec   &  SB(s)a:  & 0.42 & T/C  & 9.17\tablenotemark{a} & -21.7 $\pm$ 0.3 & 125 & 60
 & 88 & 440 $\pm$ 6 & 3.1 & 15.2 $\pm$ 1.9 \\
NGC 4429 & 12 27 26.4 &  11 06 29  &   S0$_{3}$(6)/Sa pec &SA(r)0$^{+}$ & 0.54 & T/A & 6.76 & -24.3 $\pm$ 0.1 & 220 & 60 & 89 & 1127 $\pm$ 31 & 1.5 & 16.3 $\pm$ 0.9 \\
NGC 4450 & 12 28 29.3 &  17 05 07  &   Sab pec  &  SA(s)ab   & 0.49 &   T/A  & 6.90 & -23.6 $\pm^{0.1}_{0.2}$ & 205  & 46 & 179 & 1958 $\pm$ 6 & 4.7 & 12.6 $\pm^{1.0}_{0.6}$\\
NGC 4457 & 12 28 59.3 &  03 34 16 &    RSb(rs) II & (R)SAB(s)0/A & 0.66 & T/N[s] & 7.96 & -23.1 & 160\tablenotemark{b} & 34 \tablenotemark{c} & 79\tablenotemark{d} & 881 $\pm$ 14 & 8.8 & 16\tablenotemark{e} \\
NGC 4569 & 12 36 49.8 &  13 09 46 &    Sab(s) I-II & SAB(rs)ab  & 0.43 & T/N[s] & 6.77 & -23.80 $\pm$ 0.01 & 266 & 64 & 26 & -232 $\pm$ 22 & 1.7 & 13.0 $\pm$ 0.1\tablenotemark{f}\\
NGC 4580 & 12 37 48.6 &  05 22 06  &   Sc/Sa   &    SAB(rs)a pec & 0.41 & T/N[s] & 8.77 & -22.95 $\pm$ 0.05 & 87 & 45 & 154 & 1036 $\pm$ 7 & 7.2 & 22.1 $\pm$ 0.5\\
NGC 4606 & 12 40 57.6 &  11 54 44  &   Sa pec   &   SB(s)a:  & 0.42 &  T/C & 9.30 & -22.20 $\pm^{0.05}_{0.11}$ & 103  & 67 & 44 & 1655 $\pm$ 16 & 2.5 & 19.9 $\pm^{1.0}_{0.5}$\\
NGC 4651 & 12 43 42.6 &  16 23 36  &   Sc(r) I-II & SA(rs)c  & 0.55 &  N   & 8.25 & -23.90 $\pm^{0.10}_{0.04}$ & 157 & 51 & 72 & 804 $\pm$ 10 & 5.1 & 26.9 $\pm^{0.5}_{1.2}$\\
NGC 4694 &  12 48 15.1 &  10 59 00  &   Amorph     & SB0 pec  & 0.62 & T/C   & 9.03 & -21.6 $\pm$ 0.2 & 120 & 42 & 146 & 1177 $\pm$ 11 & 4.5 & 13.4 $\pm^{1.3}_{1.0}$ \\
NGC 4698 & 12 48 23.0 &  08 29 14 &    Sa     &     SA(s)ab  & 0.51 & A  & 7.43 & -24.50 $\pm_{0.41}^{0.83}$ & 187 & 62 & 169& 1005 $\pm$ 15 & 5.8 & 24.3 $\pm^{5.1}_{4.2}$\tablenotemark{g} \\
\enddata
\tablenotetext{a}{2MASS Galaxy Atlas, Jarrett \etal 2003.}
\tablenotetext{b}{Extrapolation from R-band light profile}
\tablenotetext{c}{HyperLEDA database}
\tablenotetext{d}{Measured at 100" $\sim$ 0.6R$_{25}$}
\tablenotetext{e}{Virgo Cluster distance}
\tablenotetext{f}{Stellar kinematics based distance considering dark matter halo from Cort\'es \etal 2008}
\tablenotetext{g}{HI-based distance from Solanes \etal 2002}
\tablecomments{(1) Galaxy name; (2) Right Ascension in hours, minutes, and seconds; (3) Declination in degrees,
minutes, and seconds; (4) Hubble Types from BST, Sandage \& Tammann 1987, or Sandage \& Bedke 1994; (5)
Hubble type from RC3; (6) Central R light concentration parameter (Koopmann \etal 2001); (7) Star formation class from
Koopmann \& Kenney 2004; (8) Apparent magnitude 
in $H$ band (Gavazzi \etal 1999); (9) Absolute magnitude in $H$ band (Cort\'es \etal 2008); (10) Radius in units of arc seconds at the 25 R mag
arcsec$^{-2}$ isophote (Koopmann \etal 2001); (11) Inclination from Koopmann \etal 2001; (12) Optical P.A at R$_{25}$ (Cort\'es \etal in prep.); (13) Heliocentric
radial velocity from HyperLEDA; (14) The projected angular distance in degrees of the galaxy from M87;
(15) Line-of-sight distance (Cort\'es \etal 2008).}
\end{deluxetable}
\end{turnpage}
\clearpage
\global\pdfpageattr\expandafter{\the\pdfpageattr/Rotate 90}

\begin{deluxetable}{cccc}
\tablewidth{0pc}
\tablecaption{DensePak Observation Logs\label{table2}}
\tablehead{
\colhead{ } & \colhead{ } & \colhead{P.A.$_{\rm DensePak}$} & \colhead{T$_{\rm exp}$} \\
\colhead{Name} & \colhead{Dates} & \colhead{($\deg$)} & \colhead{(sec)} \\
\colhead{(1)} & \colhead{(2)} & \colhead{(3)} & \colhead{(4)}} 
\startdata
NGC 4604 & 8/04/1999 &   150 &  4$\times$1800 \\
NGC 4293 & 11-12/02/2002 &   72 &  1800, 2$\times$1800 \\
NGC 4351 & 25/05/2001 &   80  &  4$\times$1800 \\
NGC 4424 & 9/04/1999 &   90  &  6$\times$1800 \\
NGC 4429 &  12/02/2002 &   99 &    2$\times$1800 \\
NGC 4450 &  12/02/2002 &  175 &  4$\times$1800 \\
NGC 4457 & 11/02/2002 &    0 &    4$\times$1800 \\
NGC 4569 & 24/05/2001 &    0 &    4$\times$1800 \\
NGC 4580 & 11/02/2002 &  158  &   4$\times$1800 \\
NGC 4606 & 12/02/2002 &    33 &   4$\times$1800 \\
NGC 4651 & 24/05/2001 &   71  &  4$\times$1800  \\
NGC 4694 & 9/04/1999 & 140  &  3$\times$1800  \\
NGC 4698 & 10/02/2002 &  170  &  4$\times$1800  \\
\enddata
\tablecomments{(1) Galaxy name; (2) Observation date; (3) P.A of the DensePak array (taken from Koopmann \etal 2001); (4) Exposure time.}
\end{deluxetable}

\begin{deluxetable}{cccc}
\tablewidth{0pc}
\tablecaption{Characteristics of template stars\label{table3}}
\tablehead{
\colhead{ } & \colhead{ } & \colhead { } & \colhead{$V_{\rm helio}$} \\
\colhead{Name} & \colhead{Template Star} & \colhead{Spectral Type} & \colhead{($\kms$)} \\
\colhead{(1)} & \colhead{(2)} & \colhead{(3)} & \colhead{(4)}}
\startdata
NGC 4064 & HD 90861 & K2 III& 37.13  \\
NGC 4293 & HD 69632 & K0 III& -0.90  \\
NGC 4351 & HD 159479 &  K2 III& -24.6 \\
NGC 4424 & HD 86801 & G0 V& -14.5 \\
NGC 4429 & HD 69632 & K0 III& -0.90  \\
NGC 4450 & HD 69632 & K0 III& -0.90  \\
NGC 4457 & HD 35005 & G7 III& 25.90  \\
NGC 4569 & HD 159479 &  K2 III& -24.6 \\
NGC 4580 & HD 35005 & G7 III&  25.90  \\
NGC 4606 & HD 35005 & G7 III&  25.90  \\
NGC 4651 & HD 159479 &  K2 III& -24.6  \\
NGC 4694 & HD 90861 & K2 III& 37.13 \\
NGC 4698 & HD 77823 & K2 III& 65.90 \\
\enddata
\tablecomments{(1) Galaxy name; (2) Template star name; (3) Template star spectral type; (4) Heliocentric radial
velocity (Barbier--Brossat \etal 2000).}
\end{deluxetable}

\begin{deluxetable}{ccccccccc}
\tabletypesize{\scriptsize}
\tablewidth{0pc}
\tablecaption{Parameters and derived quantities for the stellar velocity fields\label{table4}}
\tablehead{
\colhead{ } & \colhead{$V_{\rm sys}$} & \colhead{PA$_{outer}$} & \colhead{PA$_{inter}$} & \colhead{ PA$_{\rm kin}$} & \colhead{$\Psi_{outer}$} & \colhead{$\Psi_{inter}$} & \colhead{$V_{\rm rot}$} & \colhead{$R_{\rm max}$} \\
\colhead{Name} & \colhead{($\kms$)} & \colhead{($\deg$)} & \colhead{($\deg$)} & \colhead{($\deg$)} & \colhead{($\deg$)} & \colhead{($\deg$)} &  \colhead{($\kms$)} & \colhead{(arcsec)} \\
\colhead{(1)} & \colhead{(2)} & \colhead{(3)} & \colhead{(4)} & \colhead{(5)} & \colhead{(6)} & \colhead{(7)} & \colhead{(8)} & \colhead{(9)} }
\startdata
NGC 4064  & 929 $\pm$ 3 & 148 & 152 & 129\tablenotemark{a} $\pm$ 15 & 19 $\pm$ 15 & 23 $\pm$ 15 & 72 $\pm$  7 & 25 \\
NGC 4293  & 926 $\pm$ 4  & 62 & 72 & 42 $\pm$ 8 & 20 $\pm$ 8 & 30 $\pm$ 8 & 105 $\pm$ 10 & 27\\
NGC 4351  &  2310 $\pm$ 2 & 65 & 70 & 53 $\pm$ 8 & 12 $\pm$ 8 & 17 $\pm$ 8 & 41 $\pm$  19 & 20 \\
NGC 4424  &  442 $\pm$ 4  & 88 & 96 & 87 $\pm$ 8 & 1 $\pm$ 8 & 9 $\pm$ 8 &  31 $\pm$ 4 & 25 \\
NGC 4429  & 1094 $\pm$ 6 & 89 & 97 & 87 $\pm$ 3 & 2 $\pm$ 3 & 10 $\pm$ 3 & 136 $\pm$ 9 & 25 \\
NGC 4450  &  1953 $\pm$ 2 & 178 & 173 & 166\tablenotemark{b} $\pm$ 5 & 12 $\pm$ 5 & 7 $\pm$ 5 & 114 $\pm$ 6 & 25\\
NGC 4457  &  881 $\pm$ 2 &  \nodata & 77 & 70 $\pm$ 3 & \nodata & 7 $\pm$ 3 &  83 $\pm$ 5 & 20 \\
NGC 4569  & -222 $\pm$ 6 & 26 & 19 &  32 $\pm$ 6\tablenotemark{c} & 6 $\pm$ 6 & 13 $\pm$ 6 & 103 $\pm$ 3 & 25 \\
NGC 4580  & 1031 $\pm$ 4 &  154 & 160 & 158 $\pm$ 4 & 4 $\pm$ 4 & 2 $\pm$ 4 & 103 $\pm$ 8 & 25  \\
NGC 4606  &  1636 $\pm$ 5 & 44 & 39 & 39 $\pm$ 4 & 5 $\pm$ 4 & 0 $\pm$ 4 & 55 $\pm$ 4 & 25 \\
NGC 4651  &  784 $\pm$ 2  & 72 & 79 & 82 $\pm$ 4 & 10 $\pm$ 4 & 3 $\pm$ 4 & 178 $\pm$ 8 & 28 \\
NGC 4694  & 1174 $\pm$ 22 & 146 & 142 & 146 $\pm$ 6 & 0 $\pm$ 6 & 4 $\pm$ 6 & 63 $\pm$ 2 & 26\\
NGC 4698  & 1025 $\pm$ 6  & 169 & 167 & 168\tablenotemark{d} $\pm$ 3 & 1 $\pm$ 3 & 1 $\pm$ 3 & 116 $\pm$ 10 & 25 \\
\enddata
\tablenotetext{a}{Strong non-circular motions due to central bar.}
\tablenotetext{b}{The stellar kinematic PA in NGC 4450 changes from $\sim$ 175$\deg$ in the central $\sim$10" to $\sim$160$\deg$ at r $\sim$ 15-25".
This may reflect a change from the inner bulge-dominated region to the bar-dominated region at larger radius.
There is good agreement between the stellar kinematic PA for $r<$10"
and the photometric PA of the outer galaxy.}
\tablenotetext{c}{The stellar velocity field of NGC 4569 is irregular. The SE side of the velocity map suggests a different PA than the NW side,
perhaps due to dust extinction. The SE side suggests a kinematic PA closer to the photometric PA of the outer galaxy.}
\tablenotetext{d}{The mean stellar kinematic PA given here is dominated by values for $r>$10",
so is not significantly influenced by the small orthogonally rotating bulge.}

\tablecomments{(1) Galaxy name; (2) Heliocentric systemic velocity; (3) Photometric P.A at outer radii (R$_{25}$); (4) Photometric P.A at intermediate radii (0.5 R$_{25}$); (5) Mean kinematical P.A; (6)
Kinematical misalignment with respect to PA$_{outer}$; (7) kinematical misalignment with respect to PA$_{inter}$;
(8) Stellar rotation velocity at $R_{max}$; (9) Most external point with reliable stellar
rotation velocities.
}
\end{deluxetable}

\clearpage
\begin{turnpage}
\begin{deluxetable}{ccccccccccc}
\tabletypesize{\scriptsize}
\tablewidth{0pc}
\tablecaption{Parameters and derived quantities for the ionized gas velocity fields\label{table5}}
\tablehead{
\colhead{ } & \colhead{$V_{\rm sys}$ H$\beta$} & \colhead{$V_{\rm sys}$ [\ion{O}{3}]} & \colhead{P.A$_{\rm kin}$ H$\beta$} & \colhead{$\Psi$ H$\beta$} & \colhead{ P.A$_{kin}$ [\ion{O}{3}]} & \colhead{$\Psi$ [\ion{O}{3}]} & \colhead{$V_{\rm rot}$ H$\beta$} & \colhead{$R_{\rm max}$ H$\beta$} & \colhead{ $V_{\rm rot}$ [\ion{O}{3}]} & \colhead{$R_{\rm max}$ [\ion{O}{3}]}\\ 
\colhead{Name} & \colhead{($\kms$)} & \colhead{($\kms$)} & \colhead{($\deg$)} & \colhead{($\deg$)} & \colhead{($\deg$)} & \colhead{($\deg$)} & \colhead{($\kms$)} & \colhead{(arcsec)} & \colhead{($\kms$)} & \colhead{(arcsec)}\\
\colhead{(1)} & \colhead{(2)} & \colhead{(3)} & \colhead{(4)} & \colhead{(5)} & \colhead{(6)} & \colhead{(7)} & \colhead{(8)} & \colhead{(9)} & \colhead{(10)} & \colhead{(11)}
}
\startdata
NGC 4064  &   943 $\pm$4 &   947\tablenotemark{a} $\pm$2  &  150 & \nodata & \nodata  & \nodata & 31 $\pm$ 9 & 12 & \nodata & \nodata \\
NGC 4351  &   2305 $\pm$ 3 & 2309 $\pm$ 2 & 62 $\pm$ 16 & 8 $\pm$ 16 & 62 $\pm$ 16 & 8 $\pm$ 16 & 66 $\pm$ 16 & 26 & 84 $\pm$ 6 & 28 \\
NGC 4450  &   \nodata & 1960 $\pm$ 9 & \nodata & \nodata & 194 $\pm$8 & 19 $\pm$ 8 & \nodata & \nodata & 141 $\pm$ 23 & 20 \\
NGC 4457  & 858 $\pm$ 6 & 854 $\pm$ 6 & \nodata& \nodata & \nodata & \nodata & \nodata & \nodata & \nodata & \nodata \\
NGC 4569  & -294$^{a}$ $\pm$ 3 & -240 $\pm$ 4 & \nodata & \nodata & \nodata & \nodata & \nodata & \nodata & \nodata & \nodata \\
NGC 4580  &  1029 $\pm$ 3 & \nodata  & 158 $\pm$ 4 & 0 $\pm$ 4 & \nodata & \nodata & 113 $\pm$ 10 & 24 & \nodata & \nodata \\
NGC 4606  &  1647\tablenotemark{a} $\pm$ 1 & 1650\tablenotemark{a} $\pm$ 3 & \nodata & \nodata & \nodata & \nodata & \nodata & \nodata & \nodata & \nodata \\
NGC 4651  & 778 $\pm$ 6 & 778 $\pm$ 4 & 79 $\pm$ 2 & 8 $\pm$ 2 &  76 $\pm$ 2 & 5 $\pm$ 2 & 213 $\pm$ 5 & 28 & 217 $\pm$ 10 & 28 \\
NGC 4694 &  1152 $\pm$ 2 & 1161 $\pm$ 3 & \nodata & \nodata & \nodata & \nodata & \nodata & \nodata & \nodata & \nodata \\
NGC 4698 & \nodata &  978\tablenotemark{a} $\pm$ 7 & \nodata & \nodata & \nodata & \nodata & \nodata & \nodata & 163 $\pm$ 12 & 26
\enddata
\tablenotetext{a}{Velocity at central position of the array}
\tablecomments{(1) Galaxy name; (2) H$\beta$ systemic velocity; (3) [\ion{O}{3}] systemic
Velocity; (4) H$\beta$ kinematic PA; (5) H$\beta$ Kinematic PA misalignment; (6) [\ion{O}{3}] kinematic P.A.; (7) [\ion{O}{3}] Kinematic PA misalignment;
(8) H$\beta$ rotation velocity at last
measurement; (9) Most external point with reliable H$\beta$ 
rotation velocities; (10) [\ion{O}{3}] rotation velocity at last measurement; (11) Most external point with reliable [\ion{O}{3}]
rotation velocities. 
}
\end{deluxetable}
\end{turnpage}
\clearpage
\global\pdfpageattr\expandafter{\the\pdfpageattr/Rotate 90}
\clearpage

\begin{deluxetable}{ccc}
\tablewidth{0pc}
\tablecaption{Parameters derived from the stellar velocity dispersion radial profiles\label{table6}}
\tablehead{
\colhead{} & \colhead{$\sigma_{0}$} & \colhead{$\alpha$} \\
\colhead{Name} & \colhead{($\kms$)} & \colhead{($\kms$ kpc$^{-1}$)}\\
\colhead{(1)} & \colhead{(2)} & \colhead{(3)}}
\startdata
NGC 4064 &  61 $\pm$ 1 &  -4.6 $\pm$ 2.3 \\
NGC 4293 &  94 $\pm$ 4 &  -16.0 $\pm$ 5.8 \\
NGC 4351 &  24 $\pm$ 2 &   -1.0 $\pm$ 1.0\\
NGC 4424 &  47 $\pm$ 3 &  -2.7 $\pm$ 2.7 \\
NGC 4429 &  203 $\pm$ 4 &  -12.7 $\pm$ 5.1\\
NGC 4450 &  139 $\pm$ 2 &  -36.0 $\pm$ 4.9\\
NGC 4457 &  99  $\pm$ 2 & -10.3 $\pm$ 2.6 \\
NGC 4569 &  118 $\pm$ 2 &  -25.4 $\pm$ 2.9\\
NGC 4580 &  58  $\pm$ 3 &  -5.6 $\pm$ 2.8\\
NGC 4606 &  47  $\pm$ 3 &  -3.1 $\pm$ 3.1\\
NGC 4651 &  112 $\pm$ 2 &  -22.3 $\pm$ 1.5\\
NGC 4694 &  42  $\pm$ 2 &   6.2 $\pm$ 3.1\\
NGC 4698 & 143  $\pm$ 3 &  -3.4 $\pm$ 3.4\\
\enddata
\tablecomments{(1) Galaxy name; (2) Central velocity dispersion from a linear model for the dispersion profile; (3) Velocity dispersion
slope in $\kms$ kpc$^{-1}$.}
\end{deluxetable}

\begin{deluxetable}{cccccccccc}
\tablewidth{0pc}
\tablecaption{ Measured and modeled parameters: $\sqrt{\langle V^{2}/\sigma^{2}\rangle}$, $\lambda_{\rm R}$, and ellipticity.\label{table7}}
\tablehead{
\colhead{Name} & \colhead{$\sqrt{\langle V^{2}/\sigma^{2}\rangle}$} &  \colhead{$\lambda_{\rm R}$} & \colhead{$\epsilon$} & \colhead{R$_{\rm 1/2}$} & \colhead{$\lambda_{\rm R}{ \rm  (iso)}$} & \colhead{$\lambda_{\rm R_{\rm 1/2}}{ \rm (iso)}$} & \colhead{$k$} & \colhead{$\lambda_{\rm R_{\rm 1/2}}$} & \colhead{$\langle \epsilon\rangle_{\rm 1/2}$}  \\
\colhead{(1)} & \colhead{(2)} & \colhead{(3)} & \colhead{(4)} & \colhead{(5)} & \colhead{(6)} & \colhead{(7)} & \colhead{(8)} & \colhead{(9)} & \colhead{(10)} }
\startdata
NGC 4064 & 0.61 $\pm$ 0.05 & 0.56 & 0.69 & 37 & 0.69 & 0.75 & 0.81 & 0.61 & 0.63 \\
NGC 4293 & 0.47 $\pm$ 0.08 & 0.44 & 0.39 & 67 & 0.76 & 0.91 & 0.58 & 0.53 & 0.66 \\
NGC 4424 & 0.37 $\pm$ 0.09 & 0.36 & 0.61 & 39 & 0.79 & 0.82 & 0.46 & 0.37 & 0.66 \\
NGC 4429 & 0.41 $\pm$ 0.02 & 0.38 & 0.41 & 61 & 0.63 & 0.85 & 0.60 & 0.51 & 0.56 \\
NGC 4450 & 0.35 $\pm$ 0.03 & 0.40 & 0.26 & 53 & 0.72 & 0.82 & 0.56 & 0.46 & 0.35 \\
NGC 4457 & 0.31 $\pm$ 0.08 & 0.28 & 0.20 & 26 & 0.61 & 0.62 & 0.46 & 0.29 & 0.16 \\
NGC 4569 & 0.43 $\pm$ 0.03 & 0.44 & 0.36 & 95 & 0.66 & 0.89 & 0.66 & 0.59 & 0.59 \\
NGC 4580 & 0.76 $\pm$ 0.11 & 0.62 & 0.29 & 29 & 0.70 & 0.73 & 0.89 & 0.65 & 0.30 \\
NGC 4606 & 0.74 $\pm$ 0.08 & 0.52 & 0.48 & 34 & 0.63 & 0.67 & 0.83 & 0.56 & 0.57 \\
NGC 4651 & 0.94 $\pm$ 0.04 & 0.71 & 0.34 & 36 & 0.60 & 0.68 & 1.18 & 0.80 & 0.36 \\
NGC 4694 & 0.68 $\pm$ 0.08 & 0.49 & 0.54 & 27 & 0.90 & 0.90 & 0.55 & 0.49 & 0.48 \\
NGC 4698 & 0.28 $\pm$ 0.03 & 0.26 & 0.07 & \nodata & \nodata & \nodata & \nodata & \nodata & \nodata \\
\enddata
\tablecomments{
(1) Galaxy name; (2) Luminosity-weighted $V/\sigma$;
(3) $\lambda_{\rm R}$ measured over the entire array (25"); (4) Ellipticity estimated within the inner 25" from 2MASS H-band images (Jarrett \etal 2003);
(5) Half-light radius in arcsec; (6) $\lambda_{\rm R iso}$ over the inner 25" derived from isotropic two-integral dynamical models (Cort\'es \etal 2008);
(7) $\lambda_{\rm R_{\rm 1/2} iso}$ over the inner R$_{\rm 1/2}$ derived from isotropic two-integral dynamical models (Cort\'es \etal 2008);
(8) $k$, ratio between observed $\lambda_{\rm R}$ measured
over the inner 25" to $\lambda_{\rm R}$ within the inner 25" derived from isotropic two-integral dynamical models; (9) $\lambda_{\rm R}$ estimated in the inner R$_{\rm 1/2}$
from two-integral dynamical models corrected by isotropy; (10) luminosity-weighted ellipticity within the inner R$_{1/2}$.}

\end{deluxetable}

\clearpage
\begin{turnpage}
\begin{deluxetable}{clll}
\tabletypesize{\scriptsize}
\tablewidth{0pt}
\tablecaption{Kinematic disturbances \& evidence of Interaction\label{table8}}
\tablehead{
\colhead{Galaxy} & \colhead{Morphological Features} & \colhead{Kinematic Features} & \colhead{Interaction} \\
\colhead{(1)} & \colhead{(2)} & \colhead{(3)} & \colhead{(4)} }
\startdata
NGC 4064 & very strong stellar bar & very strong bar streaming motions & gravitational interaction \\
         & disturbed dust morphology & flat $\sigma$ profile  & ram pressure stripping\\
         & truncated gas disk &  & \\
         & [T/C] H$\alpha$ morphology & & \\
         &  & & \\
NGC 4293 &  outer stellar disk warped &  kinematic misalignment \& twist & gravitational interaction \\
         & disturbed dust lanes &  & \\
         &  & & \\
NGC 4351 &  lopsided & curved isovelocity contours for star \& gas & ongoing ram pressure stripping?\\
         & HI compressed in NE \& extended to SW & stars and gas misaligned &  \\
         &  & & \\
NGC 4424 &  heart-shaped stellar body \& shells &  low $v/\sigma$ and flat $\sigma$ profile & recent intermediate-mass ratio merger\\
         & large HI tail & off-center double $\sigma$ peaks & or close high-velocity collision\\
         & [T/C] H$\alpha$ morphology & complex central gas motions & ram pressure stripping?  \\
         &  & & \\
NGC 4429 &  central dust disk & stellar $V$ rises rapidly in center then falls & old merger with gas infall to center\\
         &  cold circumnuclear stellar disk & pinched stellar isovelocity contours &  \\
         &      & $v$--$h_{3}$  anti-correlation &  \\
         &      & central $\sigma$ drop & \\
         &      & LOSVD tail & \\
         &      & misalignment between central disk and outer body & \\
         &  & & \\
NGC 4450 &  cold circumnuclear stellar disk &  stellar \& [\ion{O}{3}] kinematic PA misalignment & gas accretion or minor merger \\
         &      & $v$--$h_{3}$  anti-correlation &  \\
         &      & stellar V rises rapidly in center then falls & \\
         &      & pinched stellar isovelocity contours & \\
         &      & LOSVD tail & \\
         &  & & \\
NGC 4457 &  peculiar 1-arm spiral in center & curved gas isovelocity contours & ongoing ram pressure stripping? \\
         &     & gas velocities offset from stars by 30-40 km/s &   \\
         &  & & \\
NGC 4569 &  outer stellar disk warped  & non-circular gas motions in center  & post-peak active ram pressure \\
         &  with respect to inner disk & due to nuclear outflow &  \\
         &  truncated HI \& H$\alpha$ disk & misalignment between the optical & gravitational interaction \\
         &  stripped outer disk    & and kinematicsl major axes & \\
         & anomalous HI \& H$\alpha$ extraplanar arm &  & nuclear gas outflow \\
         &  & & \\
NGC 4580 &  undisturbed stellar disk &  high $v/\sigma$ & past ram pressure stripped outer disk\\
         & truncated gas disk & undisturbed gas velocity field & \\
         &  & & \\
NGC 4606 &  disturbed stellar disk  &  flat velocity dispersion profile & intermediate mass ratio merger\\
         & disturbed dust lanes & off-center double $\sigma$-peaks & \\
         & [T/C] H$\alpha$ morphology &  & \\
         &  & & \\
NGC 4651 & outer disk disturbed with shells  & undisturbed stellar and gas kinematics & recent minor merger\\
         & and tail in stars and HI  & within the central 3 kpc & \\
         & strong spiral arms at $R \sim$ 20" & LOSVD tail at $R=$10-20" & \\
         &      & strong h$_3$ component with $v$--$h_{3}$  anti-correlation & \\
         &  & & \\
NGC 4694 & complex HI distribution with tail  & central $\sigma$ drop & recent merger/accretion\\
         & extending to nearby dwarf  &  & \\
         & [T/C] H$\alpha$ morphology & flat velocity dispersion profile & \\
         &     & ionized gas kinematics disturbed &  \\
         &  & & \\
NGC 4698 &   & orthogonally rotating stellar bulge & old merger \\
         &   & flat velocity dispersion &  \\
         &   & stellar \& gas kinematic PA misalignment  &   \\
\enddata 
\tablecomments{(1) Galaxy name; (2) Relevant morphological features; (3) Relevant kinematic features; (4) Possible type of
interaction.}
\end{deluxetable}
\end{turnpage}
\clearpage
\global\pdfpageattr\expandafter{\the\pdfpageattr/Rotate 90}
\end{document}